\documentclass{bmcart}

\RequirePackage[authoryear]{natbib}
\usepackage{natbib}
\usepackage[utf8]{inputenc} 
\usepackage{multirow}
\usepackage{rotating}
\usepackage{pdflscape}
\usepackage{graphicx,subfigure}
\usepackage{hyperref}

\startlocaldefs
\endlocaldefs

\begin{document}

\begin{frontmatter}

\begin{fmbox}
\dochead{Research}


\title{Networks of international football: communities, evolution and globalization of the game}


\author[
   addressref={aff1},                   
   corref={aff1},                       
   email={yanglitjroc@gmail.com}   
]{\inits{YL}\fnm{Yang} \snm{Li}}
\author[
   addressref={aff1},
   email={gmateosb@ece.rochester.edu}   
]{\inits{GM}\fnm{Gonzalo} \snm{Mateos}}


\address[id=aff1]{%
  \orgname{Department of Electrical and Computer Engineering, University of Rochester}, 
 \city{Rochester},
 \postcode{14627},
  \cny{NY, US}
}



\end{fmbox}


\begin{abstractbox}

\begin{abstract} 
As the most popular sport around the globe, the game of football has recently intrigued much research interest to explore and distill useful and appealing information from the sport. Network science and graph-centric methods have been previously applied to study the importance of football players and teams. In this paper, for the first time we study the macroscopic evolution of the football society from a complex network point of view. Football game records within a time window of over a century were collected and expressed in a graph format, where participant teams are represented by graph nodes and the games between them are the graph edges. We carry out community detection and temporal analysis to reveal the dynamic features and the community structures embedded within the football network, offering the evidence of a continuously expanding football society. Spatio-temporal analysis is also implemented to unveil the temporal states that represent distinct development stages in the football history. Our analysis suggests that the evolution of the game receives considerable impact not only from major sport events, but also from multiple social and political incidents. The game of football and its evolution reflect significant historical transitions and turning points, and can provide a novel perspective for the study of the worldwide globalization process. 
\end{abstract}



\end{abstractbox}
%

\end{frontmatter}




\section*{Introduction}\label{s:intro}

\subsection*{The game of football and its global social impact}\label{ss:overview}
Football (also known as football or association football) is the most popular sport in the world, attracting billions of fans around the globe that regularly practice it and follow professional competitions. Originated in ancient China and subsequently popularized in England~\citep{historyfootball}, football has gradually evolved from a local sport enjoyed by only a few nations in the late 19th century, to a nowadays global sport spreading across the world and involving more than 200 countries.

The establishment of the football's international governing body in 1904, the Fédération Internationale de Football Association (FIFA), marks a milestone for the game of football to become an officially recognized sport, and football games either among clubs or between nations have been organized more systematically since then. FIFA is responsible for the organization of major international football tournaments, notably the FIFA World Cup held for the first time in 1930 and the FIFA Women's World Cup which commenced in 1991. It is hierarchically organized in terms of confederations (that can be closely mapped to continental regions), each of which comprises national association members. Table \ref{t:FIFA} presents FIFA's current 6 football confederations, along with the year in which each one was founded and the current number of national association members as of the year 2018. Through confederations, FIFA fosters the development of football within continental regions, and also promotes the worldwide progress of international football beyond geographical boundaries. 

\begin{table}[!h]
\caption{Football confederations in FIFA (as of 2018)}
      \begin{tabular}{cccc}
        \hline
        \\[-0.9em]
           Abbreviation & Name & Year founded & No. of members\\ \hline
           \\[-0.6em]
           CONMEBOL & Confederación Sudamericana de Fútbol & 1916 & 10\\
           \\[-0.6em]
           UEFA & Union of European Football Associations & 1954 & 55\\
           \\[-0.6em]
           AFC & Asian Football Confederation & 1954 & 46\\
           \\[-0.6em]
           CAF & Confédération Africaine de Football & 1957 & 54\\
           \\[-0.6em]
           CONCACAF &  \begin{tabular}{@{}c@{}}Confederation of North, Central American \\ and Caribbean Association Football\end{tabular}  & 1961 & 35\\        
           \\[-0.6em]
        OFC & Oceania Football Confederation & 1966 & 11\\ \hline
      \end{tabular}
      \label{t:FIFA}
\end{table}

From a social science vantage point, football serves as a conduit to connect different countries, different continents and most importantly, people with diverse backgrounds all over the world. Recent estimates show that more than half of the world's population consider themselves as association football fans~\citep{ssss}. For example, over 30 billion people (“accumulated” audience) watched the 2006 FIFA World Cup held in Germany. More than one billion fans tuned in to watch the final of the 2014 FIFA World Cup held in Brazil, which is one of the largest television audience of a single match in all sports. Football transcends the boundaries of sport, and brings together people from different parts of the world. As argued in~\citep{foer2004football}, football is much more than a game, or even a way of life. It's a perfect window into the crosscurrents of today's world, with all its joys and sorrows. Changes in the world may affect the global football landscape. For instance, the 1942 and 1946 FIFA World Cups, which Germany and Brazil sought to host, were canceled due to World War II and its aftermath. On the other hand, football itself may also induce socio-economic and political changes into the world. For example, over 20 million fans flocked to Germany during the 2006 World Cup, bringing not only a sporting success but also an economic, political and security boost~\citep{germanywc}.

Despite the growing global significance and increasing social impact of football, scientific research on the game is a recent endeavor, mostly aiming at objectively quantifying and identifying variables explaining the value of certain players, or the performance of specific teams. As a result, the scope of said studies is confined to localized entities such as individual (national) teams, clubs, or leagues. On the other hand, the global macro-structure of the football society remains rather unexplored. We contend that its study, facilitated by network analytic methods applied to contemporary and rich datasets, can offer novel insights into the sport, its global connectivity structure, development and evolution.

\subsection*{Previous work on network analysis of football data}\label{ss:previouswork}
Researchers have recently started to integrate network analytic methods into quantitative analyses of football. On the analysis of football players, in the study from~\citep{onody2004complex} of the bipartite network (with football players and clubs as nodes), the degree distribution decays exponentially. In~\citep{sargent2013evaluating}, the presence of highly-rated players is demonstrated to provide the most utility within a simulated team network. Small world property is later verified to exist within the networks of interactions (passes and crosses) between professional football players~\citep{gama2015small}. On a team-level analysis, a random walk approach is used in~\citep{ribeiro2010dynamics} to model a football league with an emphasis on predicting individual match scores. Their findings indicate that the dynamics of football tournaments can be accurately simulated using a simple probabilistic model. In~\citep{grund2012network}, the concept of network intensity is introduced, which is defined as the passing rate for a football game based on the analysis of team performance of the English Premier League teams. Results therein show that increases in network intensity lead to improved team performance, while increases in network centralization (the degree to which network positions are unequally distributed in a team) have the opposite effect. More recently, a set of network metrics such as density, heterogeneity and centralization is proposed in~\citep{clemente2015using} for offense analysis of football teams with the potential to aid coach decisions. 

The previous works mainly focus on localized micro-structures, such as the importance of certain players or overall team performance. So far, to the best of our knowledge, there have been no prior studies on the macro-structure of the international football landscape at a global scale. As graph-centric methods are introducing improvements in many other disciplines, we expect that their applications in the domain of social network analysis, especially in the scenario of football network analysis, would be feasible and beneficial to reveal the evolution path of the game, to reflect the turning points in football history, and to identify critical development stages of the football society for the past decades.

\subsection*{Aim of this study and contributions}\label{ss:contribution}
The aim of our study is to perform a network-analytic exploration of the macroscopic structure of the men's international football. To that end, we study for the first time a unique dataset including all the official national\footnote{\footnotesize Note that by 'national' or 'nation', we are referring to the countries or regions that are recognized by FIFA as individual participants of football games.} team football games, ranging all the way from historical clashes at FIFA World Cup finals to friendly games between islands on the Pacific where the game is only played at the amateur level. Nearly 40000 football game records from 1872 to 2016 were collected to support our study. Network graphs were constructed using the aforementioned data with teams as nodes and games as undirected edges. The networks in our study span a global scale, which means they comprehensively capture all football games between national teams happening across the entire world. Moreover, with access to the date of every football game, we are able to construct graphs for arbitrary time horizons. Such rich and dynamic graph data facilitates the analysis of football networks at different temporal resolutions.

We first investigate the existence of community structure in football networks. We test the strength of weak ties theory~\citep{granovetter1995getting} in football networks and validate the community structure linked by weak connections. Community detection algorithms are then applied to unveil various communities in a static graph including all games from 1872 to 2016, as well as for dynamic networks spanning 11 decades from 1901 to 2010. The evolution of community structures across decades is quite revealing of the development path and the global expansion of football, as only from data in recent decades that we begin to see good correspondence between communities identified and the confederations in Table~\ref{t:FIFA}. Consequently, it is intriguing to explore the landscape of football communities in early decades, and to investigate how connectivity (i.e., football game) patterns evolve into the structure we witness nowadays.

Descriptive network statistics of the time-varying networks are also generated, such as the number of games per year and the number of regional games, e.g. games played by European teams only, in each year. Such temporal features would help reveal the rate and directions of the expansion of football, which can be missing from general analyses on globalized markets and societies. With the aim of temporal states extraction, we advocate a graph similarity measure to group the graphs generated for each year from 1901 to 2010 into clusters. Each cluster represents one individual development stage consisting of several years in the football history, namely the temporal states. By referring to various social events in history, we manage to interpret each temporal state and identify the turning points that mark the boundary between states, further verifying the close relationship between the football landscape and the human history.

All in all, our work is the first to examine the macro-structure of football through its network representations. Through this perspective, we offer novel insights on the endogenous and exogenous factors driving the evolution of (community) structure and the globalization of the game. Our approach could also be translated to other domains where evolving patterns over the network are witnessed, with regard to either graph nodal attributes or graph topological connectivity, such as neuroimaging data, traffic data and internet of things (loT).

\section*{Data}\label{s:data}
In this section, we first briefly introduce the data used in our work. Preliminary exploratory analysis is carried out which reveals the micro-structure within the data and indicates the feasibility of network representation. We then provide the details regarding the construction of the football network, laying the foundation of the downstream graph-centric analysis.

\subsection*{Data collection and preprocessing}\label{ss:data}
The data used in this work contains historical football matches between men's national teams. The football match records were parsed from the World Football Elo Ratings website~\citep{eloratings,lasek2013predictive}, ranging from the first recorded and official football match on November 30, 1872 between Scotland and England, to a friendly game played between Martinique and Panama on April 27, 2016. The data set contains 39052 football match records in total, each of which contains necessary details of a football match, including the two participating teams, match venue, and match score, etc. See Table~\ref{t:samplerecord} for an example of a football match record.

\begin{table}[h!]
\caption{Example of a football match record}
\resizebox{\textwidth}{!} {
\begin{tabular}{cccccc}
    \hline
    \\[-0.9em]
    Date & Home Team & Guest Team & GoalsHome & GoalsGuest & Tournament \\ \hline
    \\[-0.6em]
    2014-06-24 & Uruguay & Italy & 1 & 0 & World Cup \\ \hline
    \\[-0.9em]
    Venue & Home Ranking* & Guest Ranking* & RankHome** & RankGuest** & ThirdPlace*** \\ \hline
    \\[-0.6em]
    Brazil & 1893 & 1831 & 9 & 15 & True\\ \hline
\\[-0.6em]
\multicolumn{4}{l}{\footnotesize * 'Ranking' refers to the Elorating ranking index on the match date}\\
\multicolumn{4}{l}{\footnotesize ** 'Rank' refers to the rank position of a team based on Elorating ranking on the match date}\\
\multicolumn{4}{l}{\footnotesize *** 'ThirdPlace' indicates if the game is held on neutral ground}\\
    \end{tabular}
    }
    \label{t:samplerecord}
\end{table}

Besides matches, we also collect information records of all the involved countries. Altogether 238 countries have participated in the game of football, i.e. each one of them has played at least one football game in history. Each record contains the name of the country, its geographical coordinates (latitude and longitude), continent and football confederation it belongs to. The geographical coordinates are used to mark each country on the map, and the confederation information is used to validate the clustering of countries via community detection (community structures of the football network). Table~\ref{t:country} shows an example record (of the country Austria).

\begin{table}[h!]
\caption{Example record of the country Austria}
      \begin{tabular}{ccccc}
        \hline
        \\[-0.9em]
           Name  & Latitude   & Longitude & Continent & Confederation\\ \hline
           \\[-0.6em]
        Austria & 47.52  & 14.55   & Europe & UEFA\\ \hline
      \end{tabular}
      \label{t:country}
\end{table}

The raw data contain all the necessary information about football matches and involved countries, but a few inconsistencies do exist. For example, several countries were split into smaller ones (e.g. collapse of the Soviet Union, East Germany/West Germany, Czechoslovakia, Yugoslavia, etc.). In addition, some countries joined together and participated in football matches as one representative regional team (e.g. Great Britain). In order to maintain data consistency and avoid data redundancy, we locate these anomalies and unify inconsistent data records, for example, assigning the geographical coordinates of England to Great Britain so that it can be located and marked on the map.

\subsection*{Mining frequent football relations among countries}\label{ss:miningrelation}
In a football game, two teams play against each other. It is common that these two teams may have played against each other before for multiple times. We call this the frequent football relations. To illustrate this feature and find micro-structures that would add up to construct a complete football network, we applied the Apriori algorithm~\citep{agrawal1996fast} to identify the frequent item sets (tuples of teams) in all games from year 1901 to year 2010, i.e. find teams that have played against each other for more than $\delta$ times ($\delta$ is the threshold, or the so-called minimum support~\citep{agrawal1996fast}). We set $\delta=11$, which is 10\% of the total number of years. Table~\ref{t:relation} shows some of the frequent relations identified that consist of different number of teams. For each frequent relation, we only list a few examples as illustration.

\begin{table}[h!]
\caption{Frequent football relations}
\resizebox{\textwidth}{!} {
\begin{tabular}{ ccc }
\hline
\\[-0.9em]
Frequent relation & Teams & Occurrence \\ \hline
\\[-0.6em]
\multirow{4}{*}{$C_3$} & England-Scotland-Wales & 69 \\
 & Denmark-Norway-Sweden & 61 \\
 & Brazil-Chile-Uruguay & 29 \\
 & Indonesia-Malaysia-Singapore & 21\\ \hline
 \\[-0.6em]
\multirow{3}{*}{$C_4$} & England-Scotland-Wales-Northern Ireland & 48 \\
 & Bahrain-Kuwait-Qatar-Saudi Arabia & 16 \\
 & Argentina-Brazil-Chile-Uruguay & 13 \\ \hline
 \\[-0.6em]
\multirow{2}{*}{$C_5$} 
 & Bahrain-Kuwait-Qatar-Saudi Arabia-Oman & 12 \\
 & Chile-Ecuador-Paraguay-Peru-Uruguay & 11 \\ \hline
 \\[-0.6em]
$C_6$ & Bahrain-Kuwait-Oman-Qatar-Saudi Arabia-United Arab Emirates & 11 \\\hline
\\[-0.6em]
\multicolumn{3}{l}{\footnotesize * $C_n$ stands for a relation involving $n$ teams}\\
\end{tabular}
\label{t:relation}
}
\end{table}

The largest relation at the threshold of 11 consists of 6 teams. The existence of these frequent relational structures indicates that the whole data set possesses some connectivity patterns. And from the table we can tell that frequent relations mostly exist between countries on the same continent, or countries from the same confederation. This finding suggests that modular structures exist within the football data, thus it is feasible to present the data as a network which could naturally capture the relationship (edges) between teams (nodes). 

\subsection*{Football network construction}\label{ss:graphconstruction}
The scientific study of networks, including computer networks, social networks, and biological networks, has received enormous amount of interest in the past decade~\citep{newman2010networks}. Networks possess the advantage of transforming complex data into a structural and systematic graph format for presentation and analysis. A network, or graph, is often denoted by $G = \{V,E,W\}$, where $V$ is the set of the nodes and $E$ is the set of edges between the nodes. $W$ is the set of edge weights for weighted graph. For unweighted binary graphs, the edge weights are set to be 1. In this work, the football networks are constructed in the following way.

\begin{itemize}
\item A time horizon is specified to delineate the temporal scope of the graph $G$;
\item Nodes in the vertex set $V$ correspond to teams that played at least one game in the prescribed time horizon;
\item Undirected edges in $E$ join a pair of nodes in $V$ if the corresponding teams played against each other (at least once); and
\item Edge weights in $W$ indicate the number of games played between teams in the prescribed time horizon.
\end{itemize}

\begin{figure}[h!]
\centering
  \includegraphics[width = 115mm, height = 60mm]{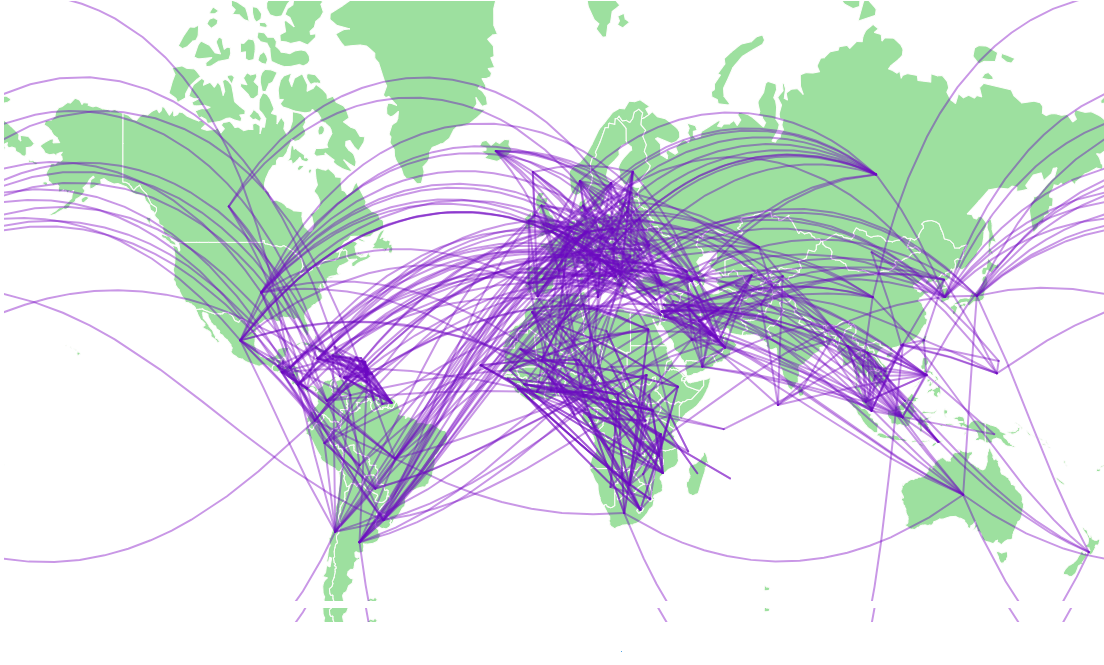} 
    \caption{Football network in 2014. The arcs (graph edges) connecting two countries (graph nodes) represent the football games played}
    \label{fig1:global}
  \end{figure}

All the constructed networks are undirected, weighted graphs. Fig.~\ref{fig1:global} shows an example of the football network constructed for the year of 2014. The arcs are the games played, and the endpoints of the edges are the participant countries, marked by their geographical coordinates on the map. 

Another way to define the edge weights is to consider the importance of the football match. As indicated by~\citep{fifarank}, different match type (World Cup, Confederation-level, Friendly, etc.) has different importance. Such importance can be integrated into the edge weights. For example, World Cup matches shall have higher weights than friendly games. In this work, we opt to use the number of games as edge weights, considering the fact that matches with higher importance are fewer in quantity compared with the total number of football matches. While integrating match importance into edge weights might be beneficial, its advantage is not clear to us. Future work shall be devoted to investigate the role of match importance in the construction of football networks.

\begin{sidewaysfigure}
  \begin{minipage}{\columnwidth}
  \centering
  \subfigure[1930]{\includegraphics[width = 0.19\textwidth, height = 0.1\textheight]{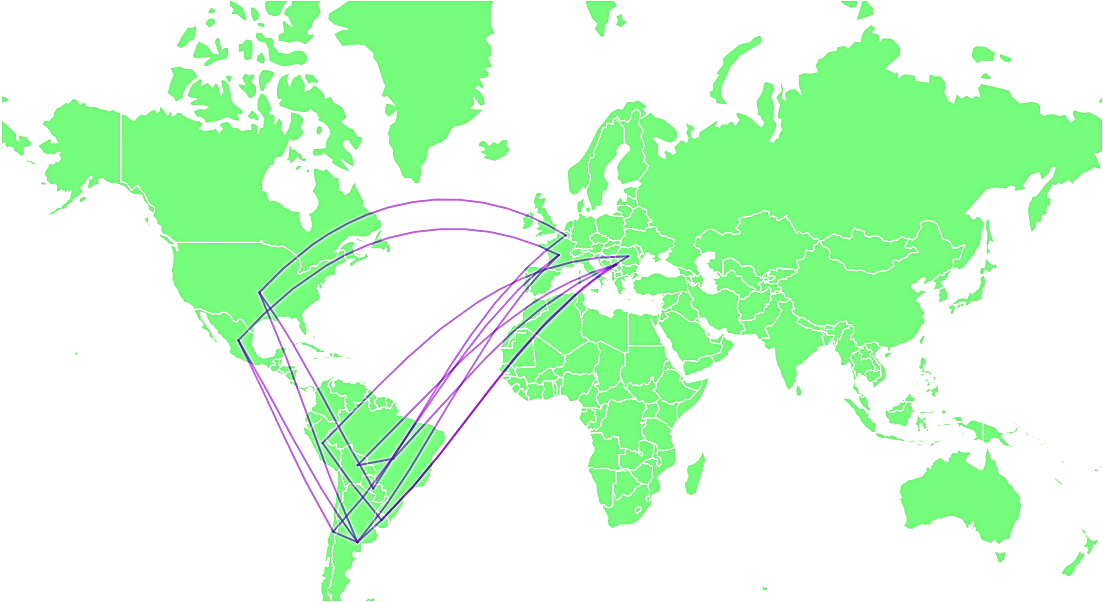}}
\subfigure[1934]{\includegraphics[width = 0.19\textwidth, height = 0.1\textheight]{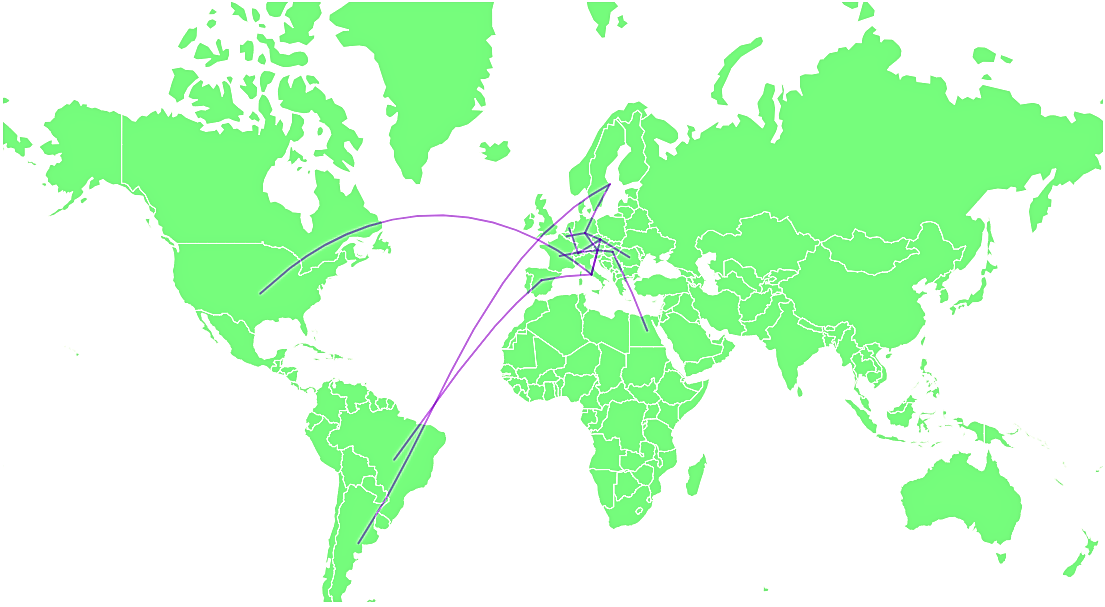}}
\subfigure[1938]{\includegraphics[width = 0.19\textwidth, height = 0.1\textheight]{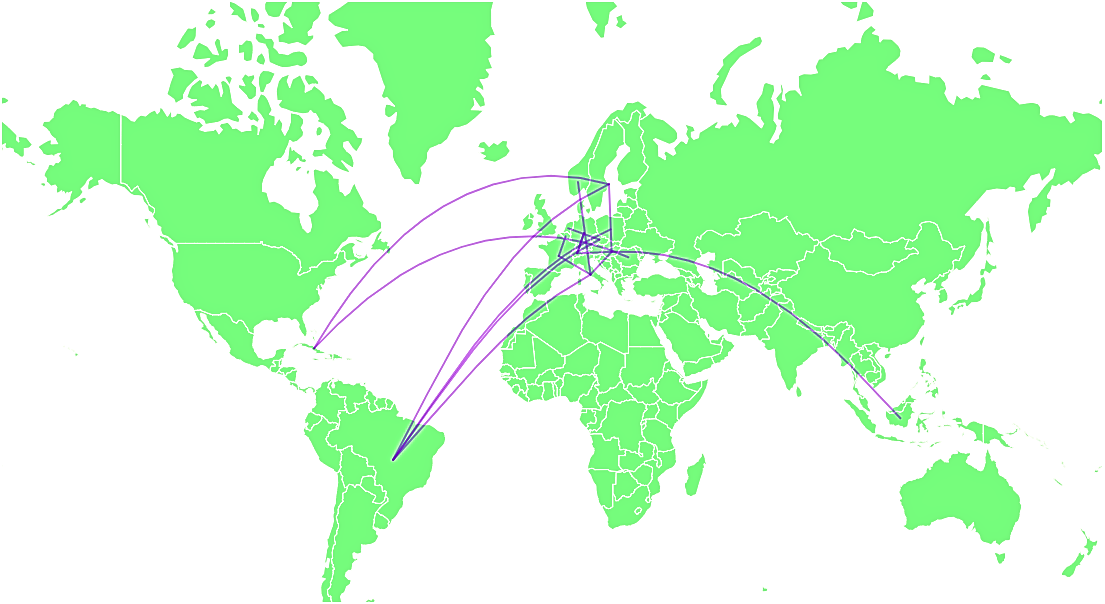}}
\subfigure[1950]{\includegraphics[width = 0.19\textwidth, height = 0.1\textheight]{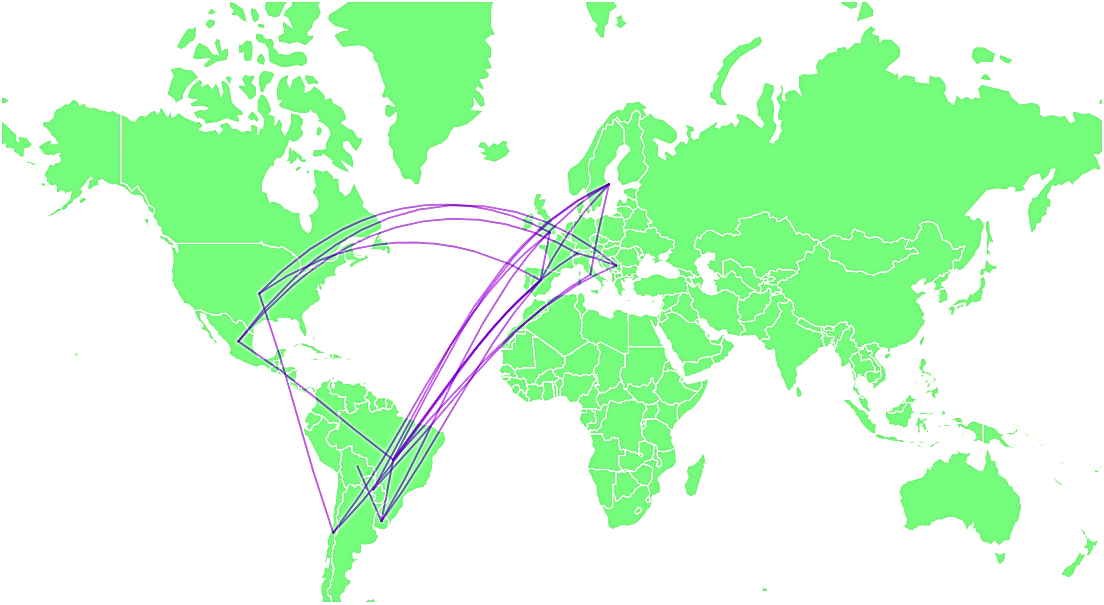}}
\subfigure[1954]{\includegraphics[width = 0.19\textwidth, height = 0.1\textheight]{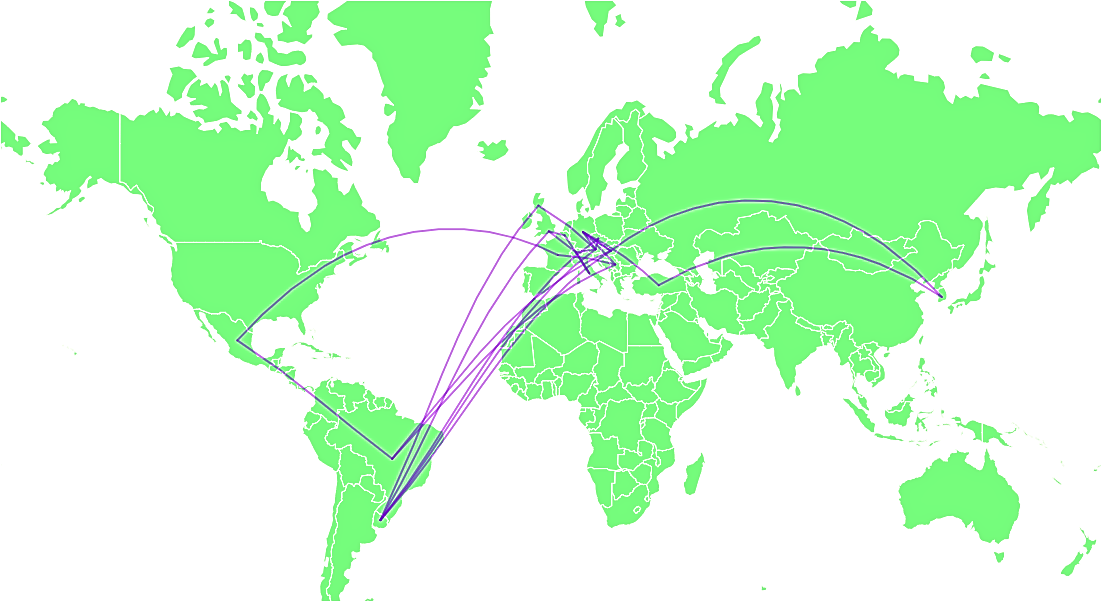}}
  \end{minipage}
  \hfill{}
  \begin{minipage}{\columnwidth}
  \centering
  \subfigure[1958]{\includegraphics[width = 0.19\textwidth, height = 0.1\textheight]{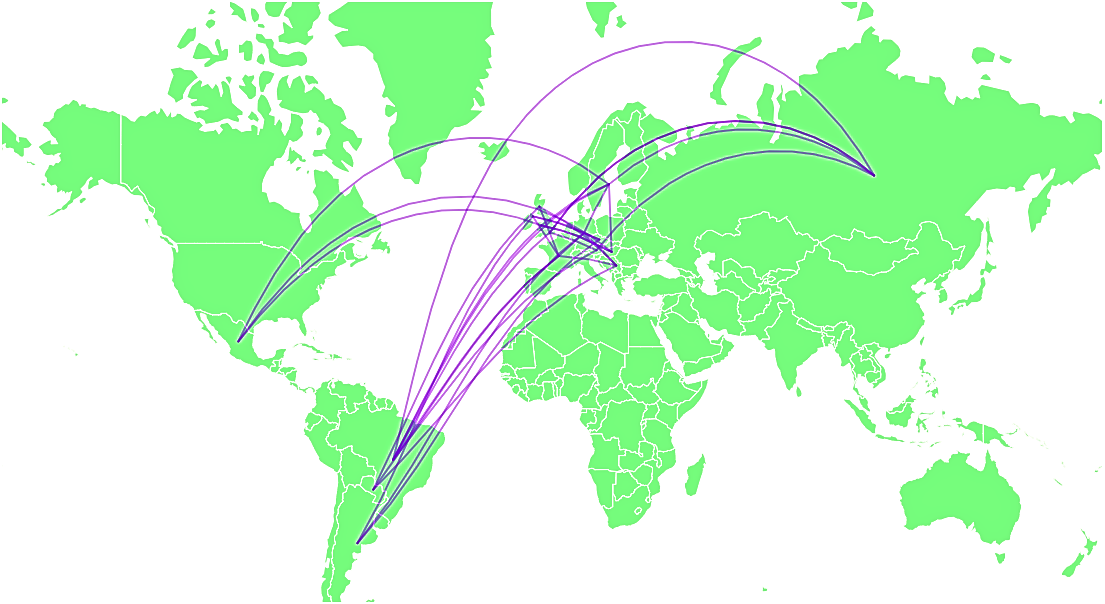}}
\subfigure[1962]{\includegraphics[width = 0.19\textwidth, height = 0.1\textheight]{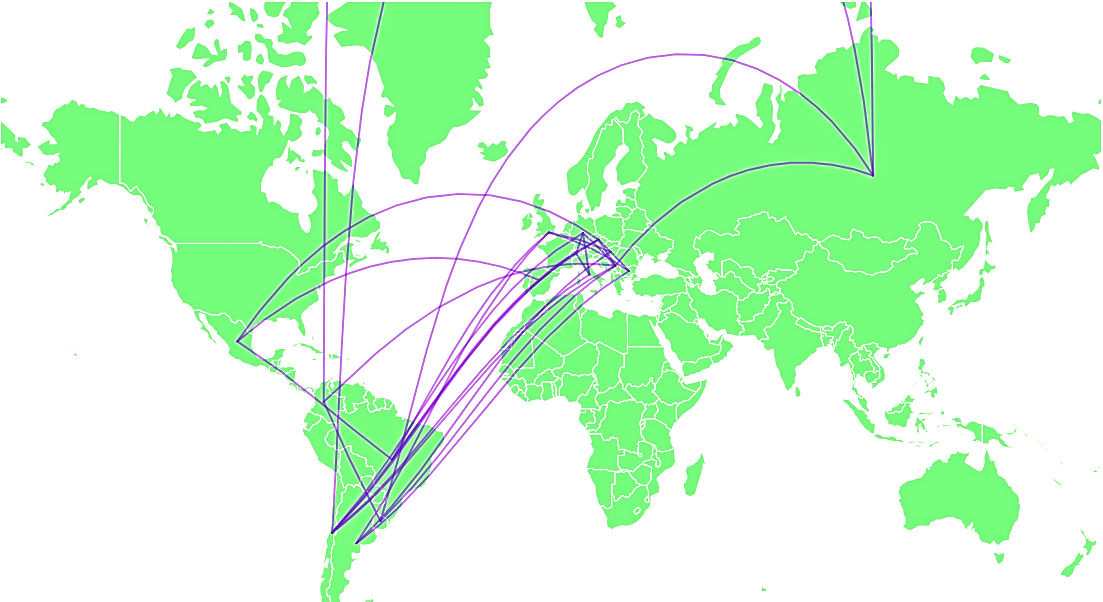}}
\subfigure[1966]{\includegraphics[width = 0.19\textwidth, height = 0.1\textheight]{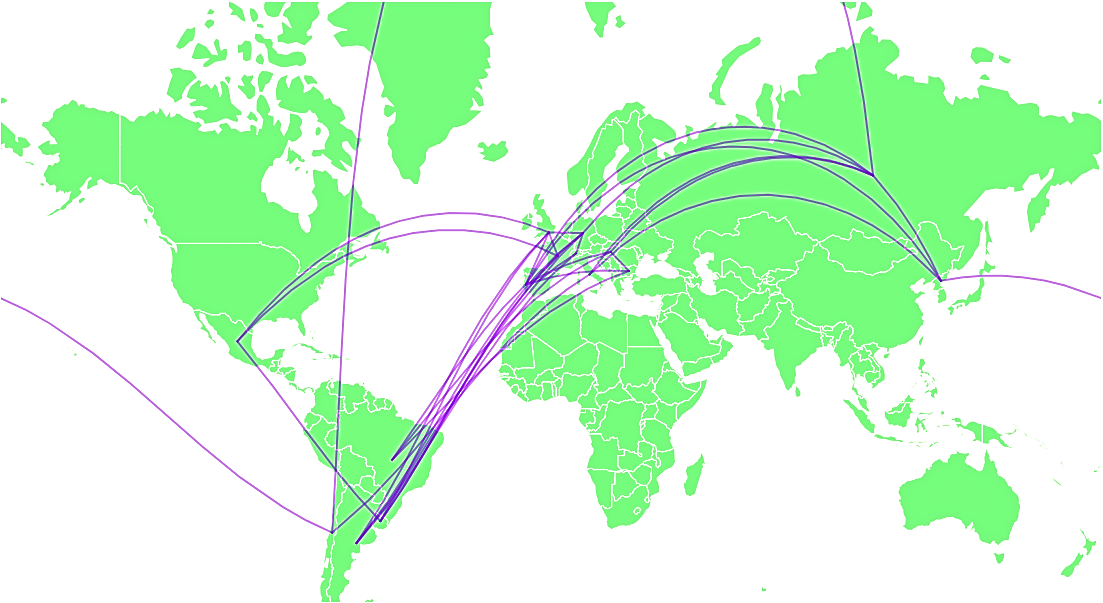}}
\subfigure[1970]{\includegraphics[width = 0.19\textwidth, height = 0.1\textheight]{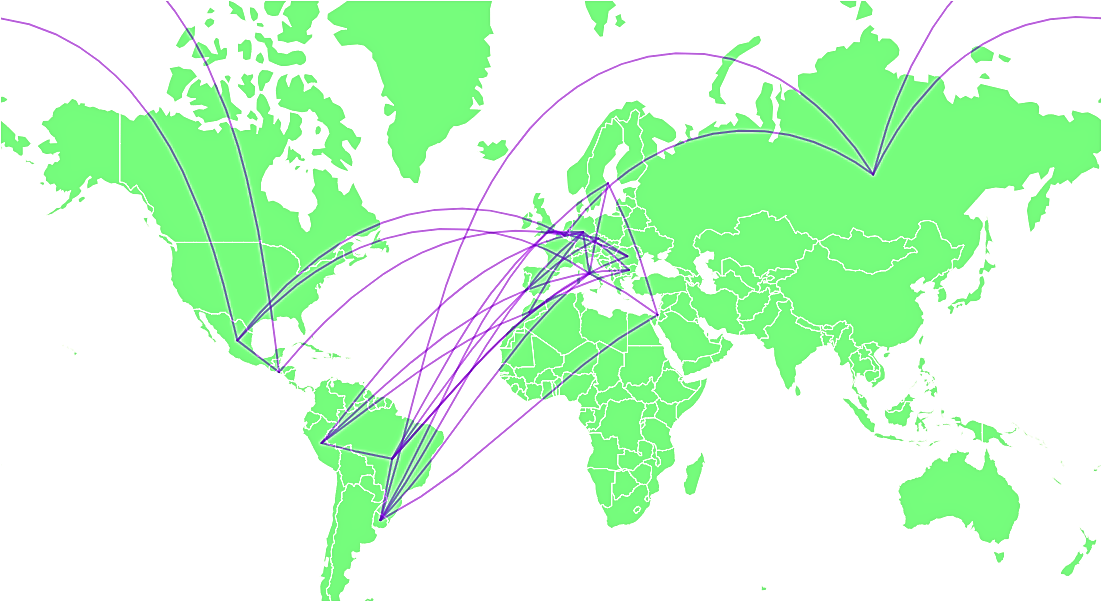}}
\subfigure[1974]{\includegraphics[width = 0.19\textwidth, height = 0.1\textheight]{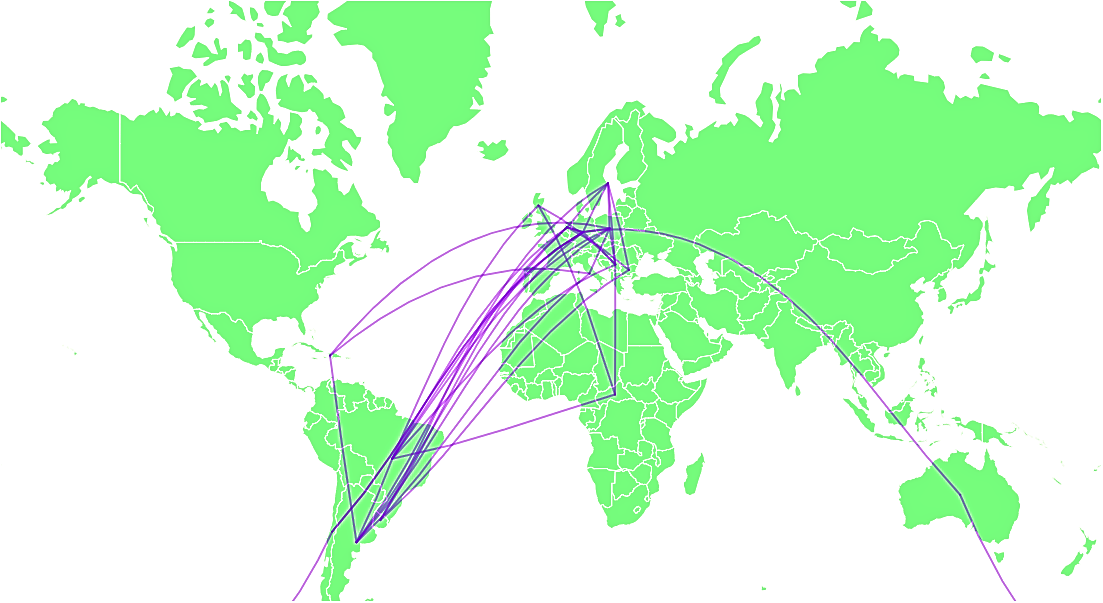}}
  \end{minipage}
  \hfill{}
  \begin{minipage}{\columnwidth}
  \centering
  \subfigure[1978]{\includegraphics[width = 0.19\textwidth, height = 0.1\textheight]{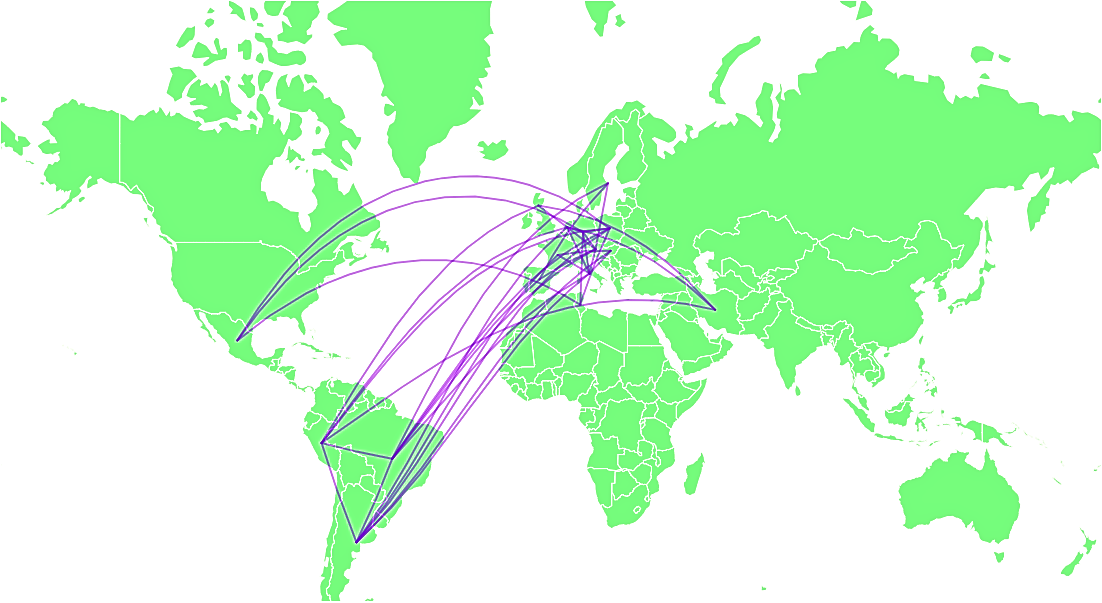}}
\subfigure[1982]{\includegraphics[width = 0.19\textwidth, height = 0.1\textheight]{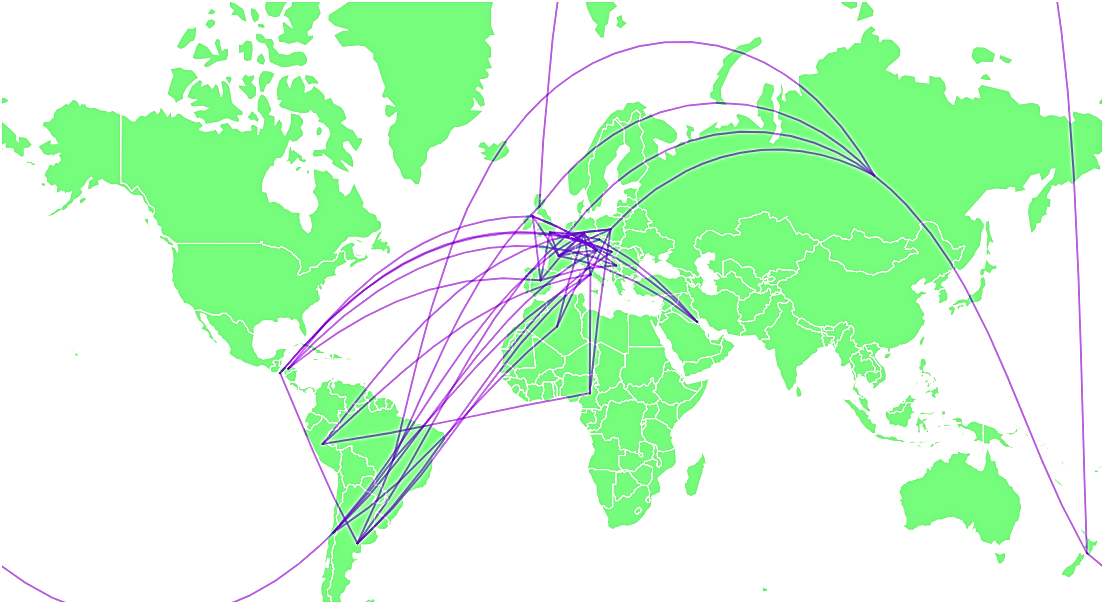}}
\subfigure[1986]{\includegraphics[width = 0.19\textwidth, height = 0.1\textheight]{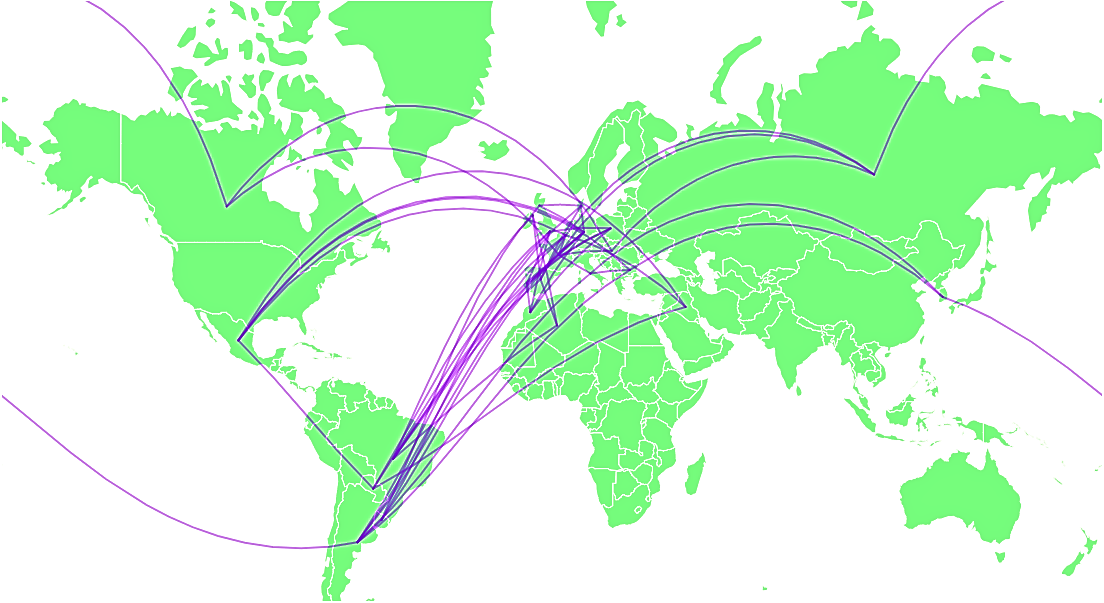}}
\subfigure[1990]{\includegraphics[width = 0.19\textwidth, height = 0.1\textheight]{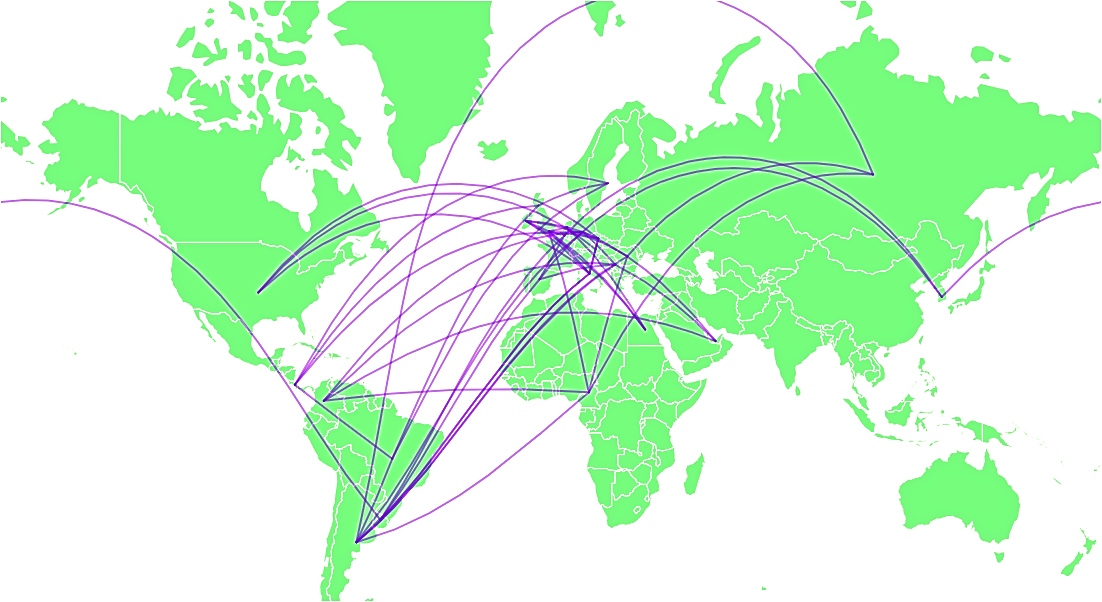}}
\subfigure[1994]{\includegraphics[width = 0.19\textwidth, height = 0.1\textheight]{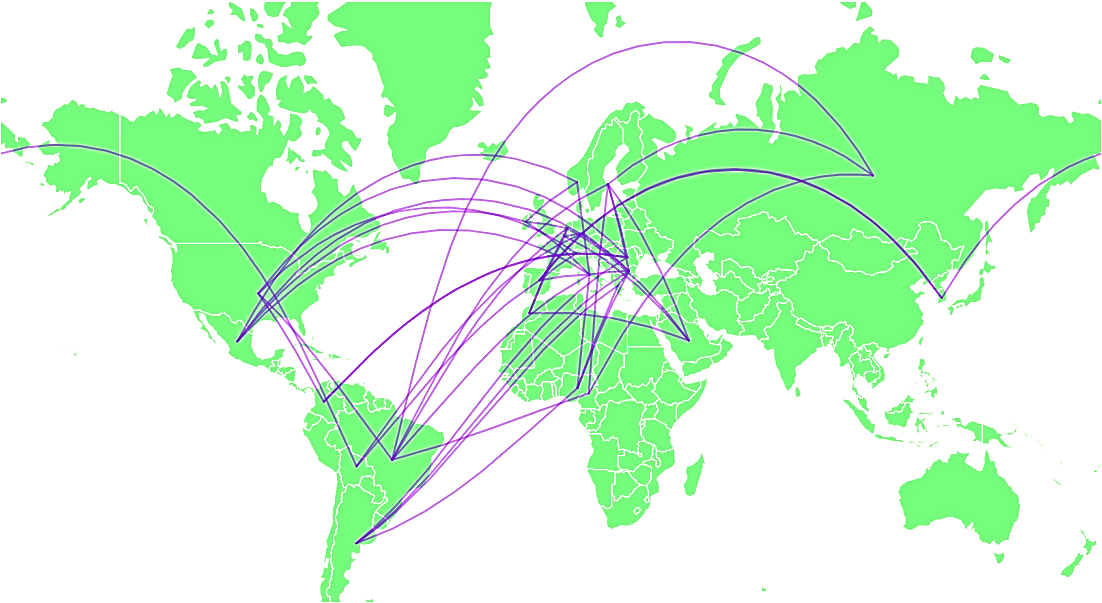}}
  \end{minipage}
  \hfill{}
  \begin{minipage}{\columnwidth}
  \centering
  \subfigure[1998]{\includegraphics[width = 0.19\textwidth, height = 0.1\textheight]{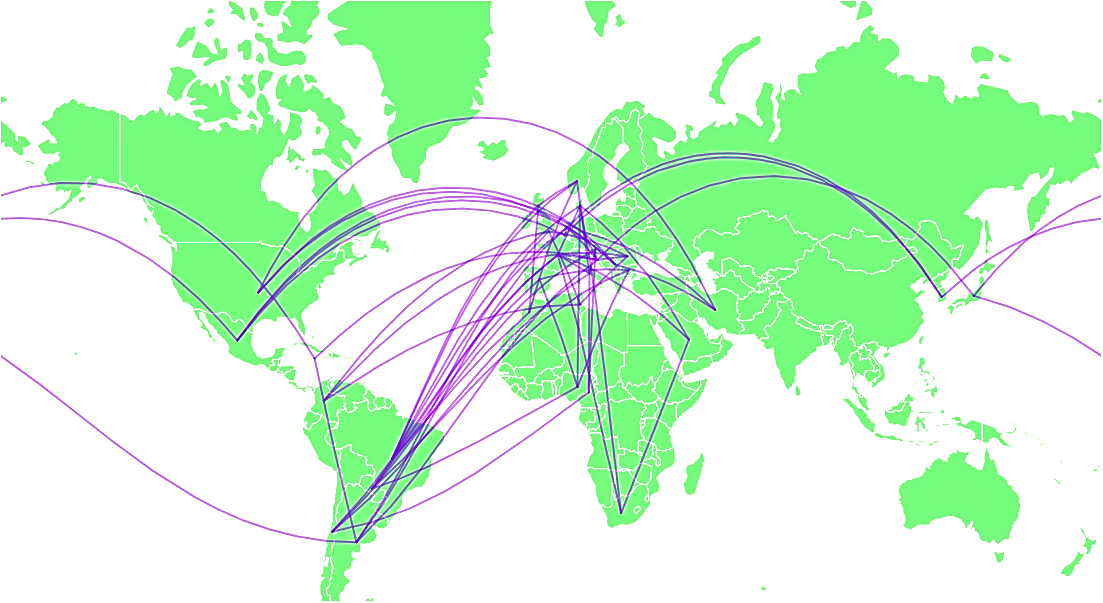}}
\subfigure[2002]{\includegraphics[width = 0.19\textwidth, height = 0.1\textheight]{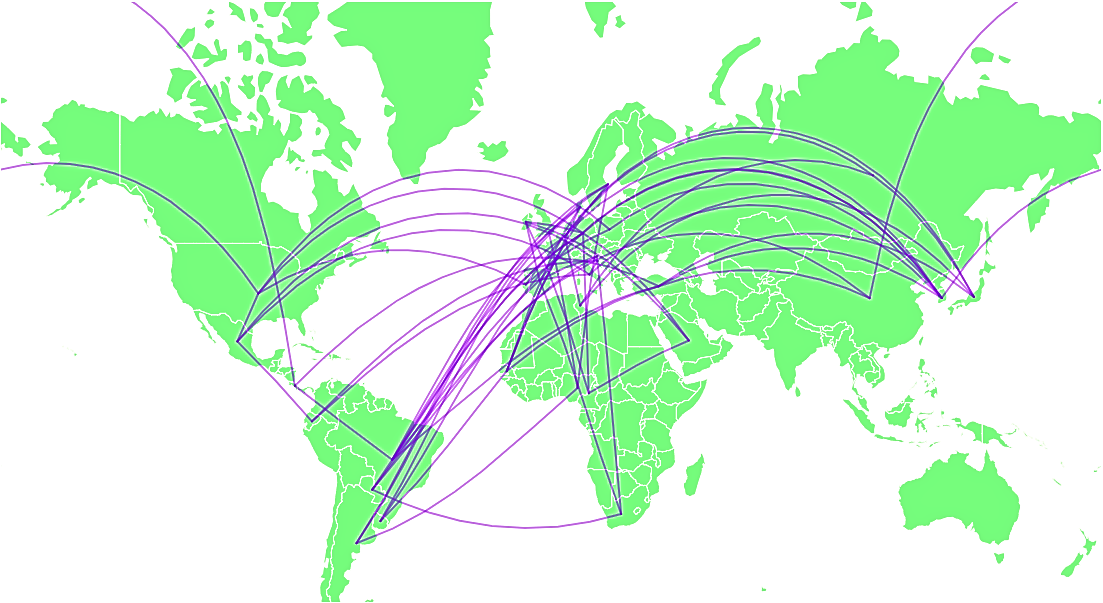}}
\subfigure[2006]{\includegraphics[width = 0.19\textwidth, height = 0.1\textheight]{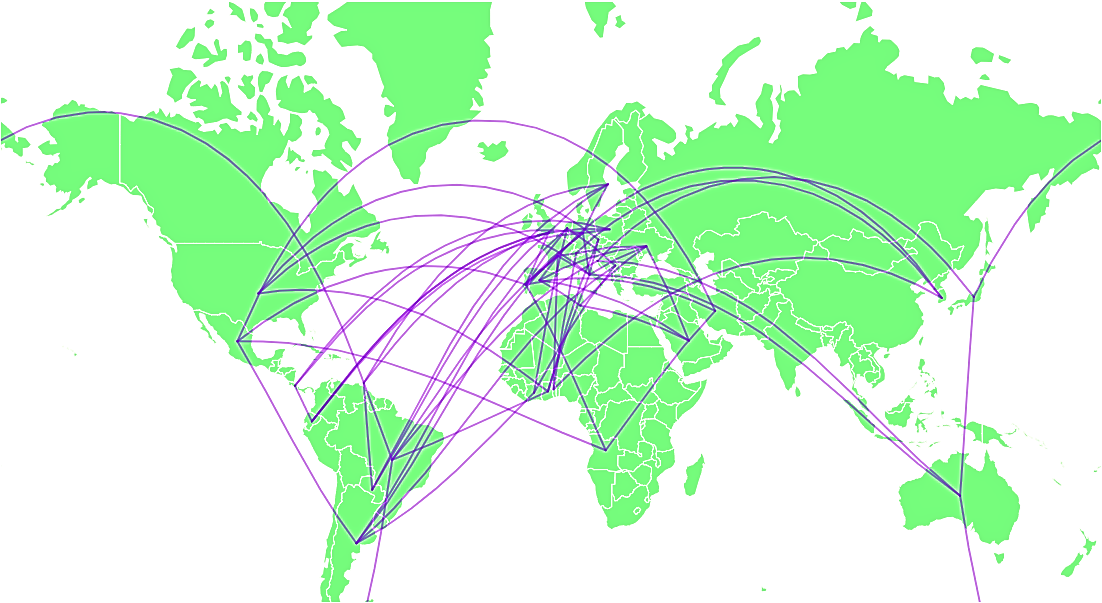}}
\subfigure[2010]{\includegraphics[width = 0.19\textwidth, height = 0.1\textheight]{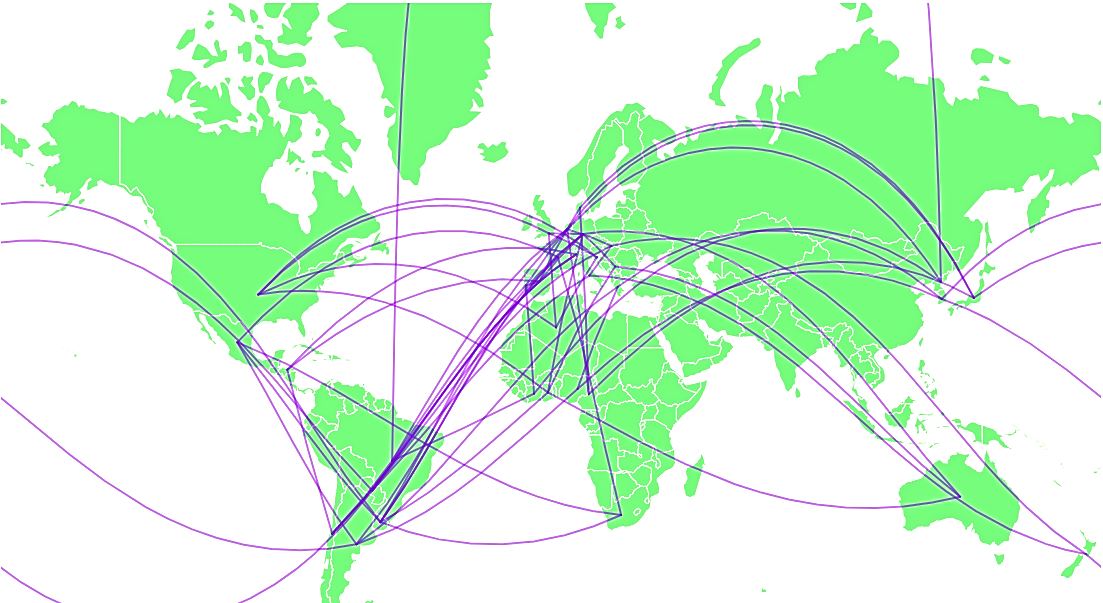}}
\subfigure[2014]{\includegraphics[width = 0.19\textwidth, height = 0.1\textheight]{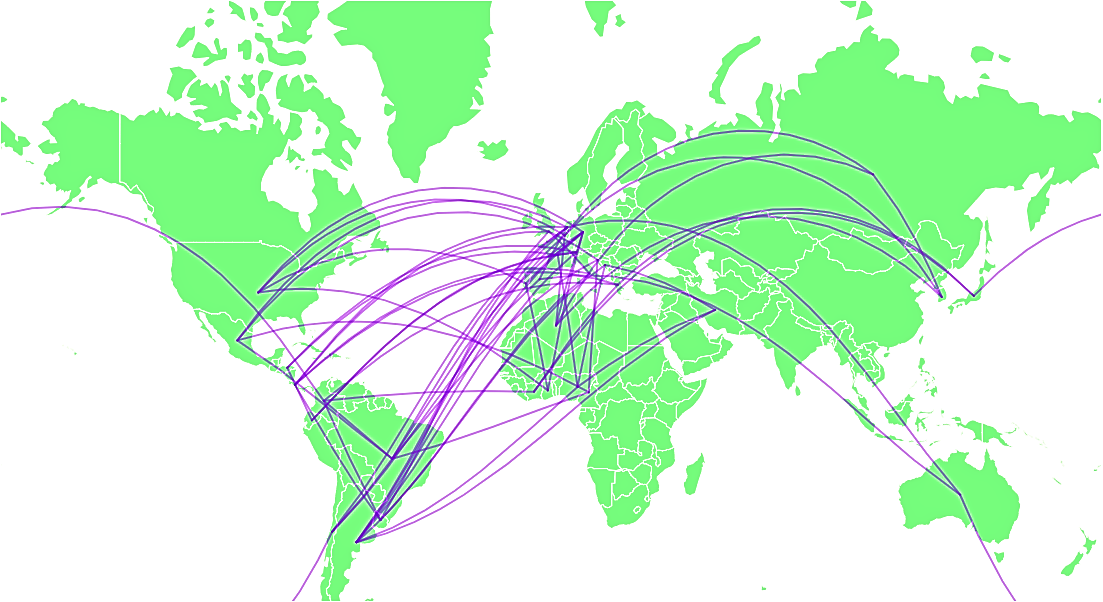}}
  \end{minipage}

 \caption{Football network for each World Cup. Each figure depicts the network of a World Cup tournament, with edges representing World Cup matches between participant countries. The first few World Cups mainly involved countries from Europe and South America. With the globalization of international football, more countries, especially Asian and African countries, get involved in the World Cup}
 \label{fig2:worldcup}
\end{sidewaysfigure}

For an example of the football networks at different timestamps, we plot in Fig.~\ref{fig2:worldcup} the football networks generated for each World Cup from 1930 to 2014. In each network, edges stand for the games played between participant countries which are located on the map using their geographical coordinates. From Fig.~\ref{fig2:worldcup}, we can clearly witness the expanding scale of the World Cup with more countries from various continents getting involved, indicating that informative temporal patterns at different timestamps do exist in this data set. These findings motivate us to exploit graph-centric methods to investigate the data, explore the information within the football network of each year, and seek to discover the temporal relationships embedded in the football history represented by a sequential series of football networks.
 
\section*{Analysis and results}\label{s:analysis}
In this section, we first briefly review the theoretical background of community detection. Then, the existence of communities within the global football network is verified by checking the extent to which Granovetter's strength of weak ties theory holds in the constructed football graphs. Descriptive graph measures are used to quantify the structural properties of the football networks, which further reveal the dynamics of the network evolution. In the end, we advocate a graph similarity measure that comprehensively integrate various graph properties, further enabling the identification of several temporal states that mark specific development stages of the football history. Via thorough comparison with social history, we manage to determine the great correspondence between the football development stages and significant events occurred in history.

\subsection*{Community detection}\label{ss:communitydetection}
One of the most important questions in social network analysis is the identification of “communities”, which are loosely defined as collections of individuals who interact unusually frequently~\citep{tantipathananandh2007framework}. Community detection aims to detect the community structure inside graphs, to identify graph modules and possibly, their hierarchical organization, by using only the information encoded in the graph topology~\citep{fortunato2010community}. Up to now, abundant methods have been proposed for community detection, and most methods can be categorized into traditional methods, modularity-based methods and others; see a thorough review of available algorithms in~\citep{fortunato2010community}.

The well-known Girvan and Newman method~\citep{girvan2002community,newman2004finding} gives a new perspective for community detection by introducing the concept of edge betweenness. Communities are detected by sequentially removing the most important edges in the network. The algorithm also introduces the term of modularity, which serves as a criterion for measuring the quality of the division of networks. The basic idea is to maximize the modularity~\citep{newman2006modularity} of the network

\begin{equation}\label{E:modularity}
	Q = \frac{1}{2|E|}\displaystyle\sum_{ij}\left[A_{ij}-\frac{k_ik_j}{2|E|}\right]\frac{s_is_j+1}{2},
\end{equation}

\noindent where $|E|$ is the total number of edges, $A_{ij}$ is the entry of the adjacency matrix on the \textit{i}th row and \textit{j}th column that connects node $i$ and $j$. $k_i,k_j$ are the degree of node $i,j$ respectively, and $s_i,s_j$ are the community assignment for node $i$ and $j$. When node $i$ and $j$ are in the same community, $s_is_j = 1$, otherwise $s_is_j = -1$.

Based on modularity optimization, a whole new set of methods has been proposed. Two advanced approaches were brought up later to speed up the detection process, often referred to as the Fast Newman's algorithm~\citep{newman2004fast} and Louvain algorithm~\citep{blondel2008fast}. In~\citep{blondel2008fast}, the algorithm first looks for communities in a local neighborhood of the node. Next, each identified community is aggregated into a new node, adding up to a new network building upon the previous one. Optimize modularity on this secondary network and repeat the steps until a maximum modularity is obtained. This method is among the fastest community detection methods. Consequently, it is implemented in this work for community detection on football networks. 

\subsection*{Strength of weak ties in football networks}\label{ss:weaktie}
The natural property of the network structure reflects its capability to bridge the local and the global components. Complex networks often optimize the tie strengths (connection between nodes) to maximize the overall flow in the network~\citep{goh2001universal,maritan1996universality}. The weak tie hypothesis
\citep{granovetter1995getting,csermely2006weak} emphasizes the importance of weak ties in connecting communities. Connections with high tie strength are more likely to be structurally-embedded within communities, whereas connections with low tie strength correlate with long-range edges joining communities. 

To verify the weak tie hypothesis and identify the intrinsic community structures of the football network, we extract a single graph including all the football games spanning from 1995 to 2015, and use participant teams as nodes and games as edges. In this graph, the numeric tie strength (i.e. edge weight) between two nodes is quantified by the total number of football games played between them. Additionally, follow the definition of the neighborhood overlap of an edge $e_{ij} \in E$ in~\citep{onnela2007structure}

\begin{equation}\label{E:neioverlap}
	O_{ij}=\frac{|n(i)\cap n(j)|}{|n(i)\cup n(j)|}=\frac{n_{ij}}{(k_i-1+k_j-1-n_{ij})},n(i):={j \in V: (i,j) \in E},
\end{equation}

\noindent where $n(i)$ is the one-hop neighborhood of the node $i$. $n_{ij}$ is the number of common neighbors shared by node $i$ and node $j$, and $k_i,k_j$ denote the degrees of node $i,j$, respectively. Edges with low overlap are related with two end nodes that do not share many common neighbors, and such edges are more likely to exist between nodes in different communities.

\begin{figure}[!t]
\centering
\subfigure[Tie strength]{\includegraphics[scale = 0.22]{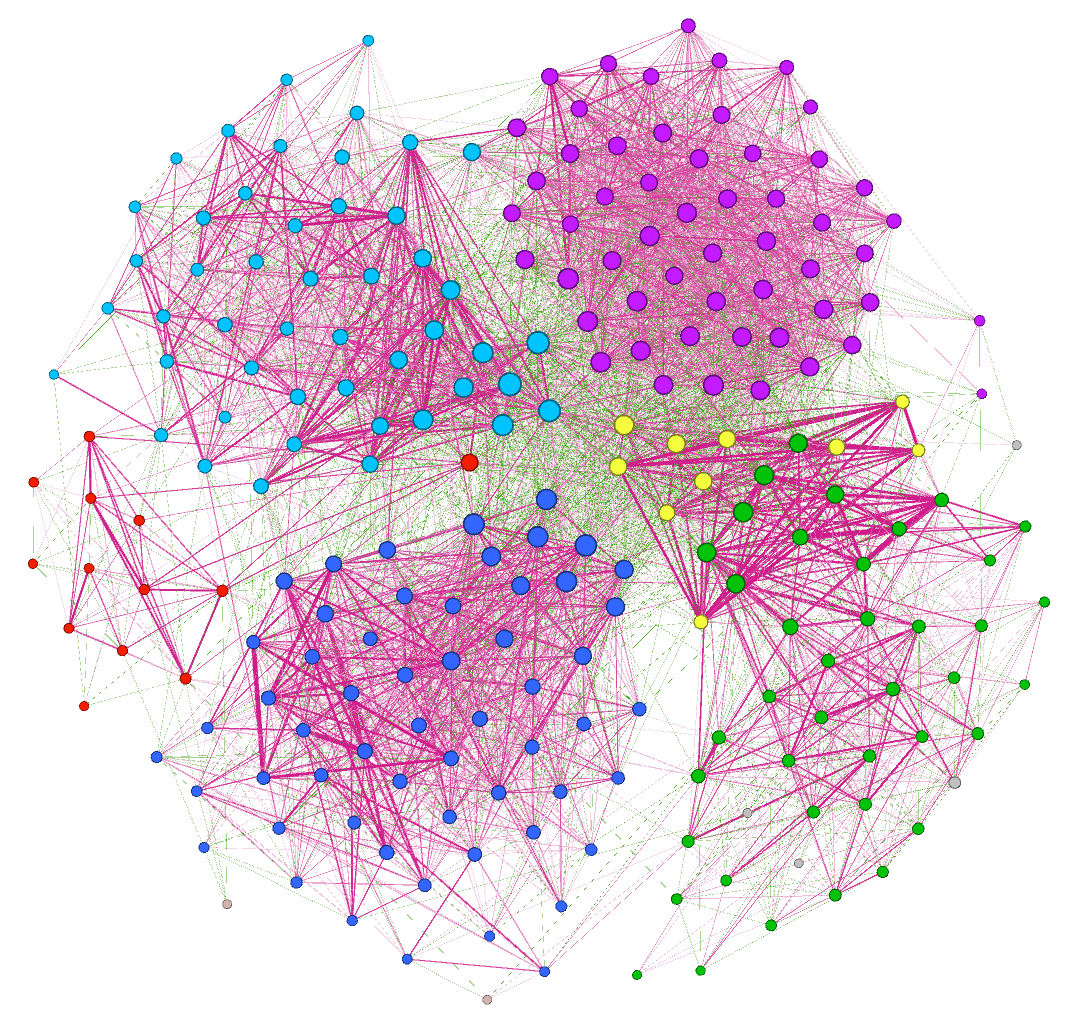}%
\label{fig3:a}}
\hfil
\subfigure[Edge overlap]{\includegraphics[scale = 0.22]{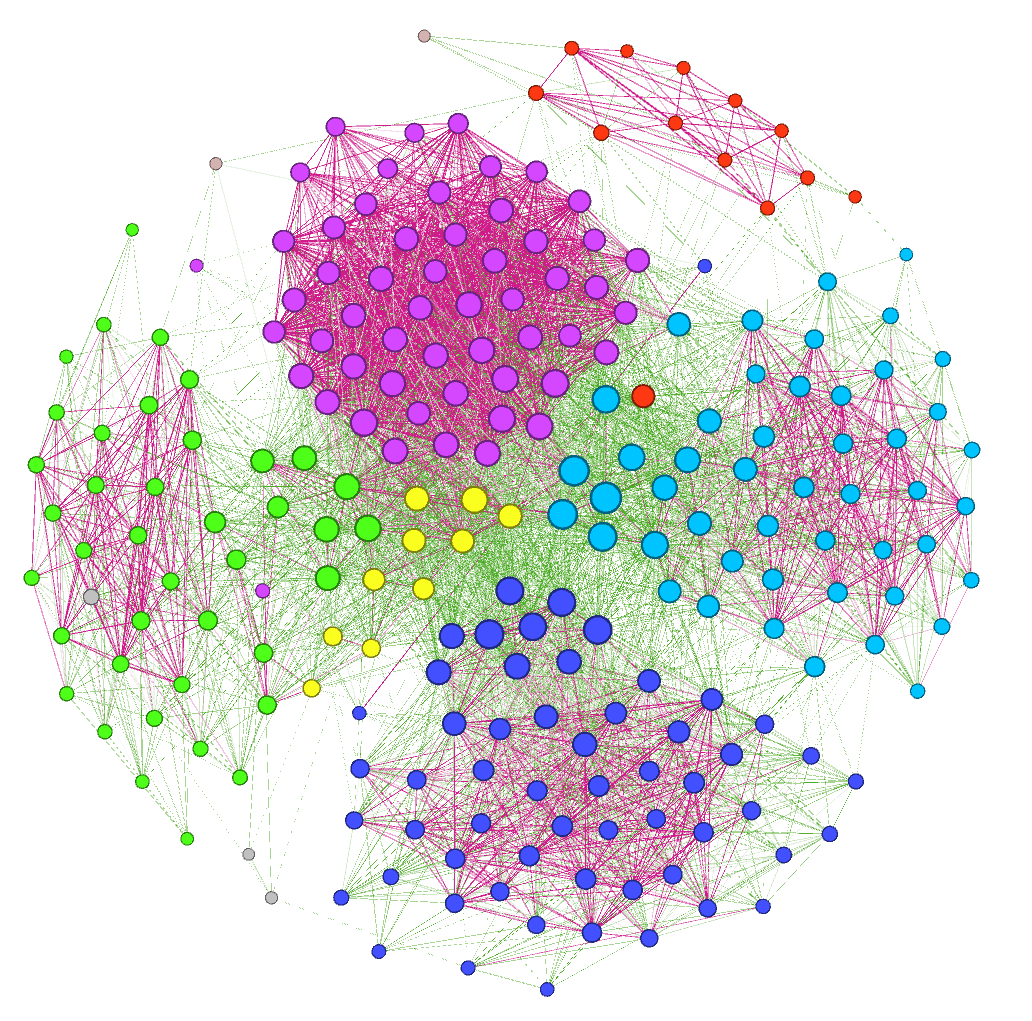}%
\label{fig3:b}}
\caption{Football network from 1995 to 2015 visualized via~\citep{bastian2009gephi}. Edges with high tie strengths (total number of football games between two countries) or edge overlap (ratio of common neighbors of the two nodes connected by the edge) are colored in red while edges with low strengths or overlap are depicted in green. Both (a) and (b) show that most green edges are between confederations and red edges are within confederations. Each node color corresponds to one confederation. Purple: UEFA; Blue: CAF; Sky blue: AFC; Red: OFC; Green: CONCACAF; Yellow: CONMEBOL}
\label{fig3:edgeoverlap}
\end{figure}

Fig.~\ref{fig3:a} and~\ref{fig3:b} show the network from year 1995 to 2015. The edge colors represent the tie strengths (edge weights) in Fig.~\ref{fig3:a} and edge overlap $O_{ij}$ in Fig.~\ref{fig3:b}, respectively. The colors of the nodes correspond to the football confederations they belong to. From the figures, it is clear that edges (in green) with low tie strengths and low overlap are mostly between confederations, while edges (in red) with high tie strengths and high overlap are mostly within confederations. To quantitatively illustrate this property, from all the 5105 edges, edges with the highest 1000 tie strength and edges with the lowest 1000 tie strength are extracted. In each group of 1000 edges, the fraction of edges connecting countries in different confederations is computed. The same procedure is also applied for edge overlap. Table~\ref{t:fraction} presents the results, which show that edges with high tie strength or high overlap are very unlikely to exist between confederations, while edges with low tie strength or low overlap are more likely to connect countries in different confederations. This result matches the weak tie hypothesis discussed earlier in this section and the visualization in Fig.~\ref{fig3:edgeoverlap}.

\begin{table}[h!]
\caption{Fractions of edges between confederations}
      \begin{tabular}{ccc}
        \hline
        \\[-0.9em]
            & Highest 1000 & Lowest 1000\\ \hline
           \\[-0.6em]
           Tie Strength & 2.3\% & 72.2\%\\
           \\[-0.6em]
           Edge Overlap & 6.6\% & 66.1\%\\
            \hline
      \end{tabular}
      \label{t:fraction}
\end{table}

\begin{figure}[h!]
\centering     
\subfigure[Tie strength vs. Edge overlap]{\label{fig4:a}\includegraphics[width = 90mm]{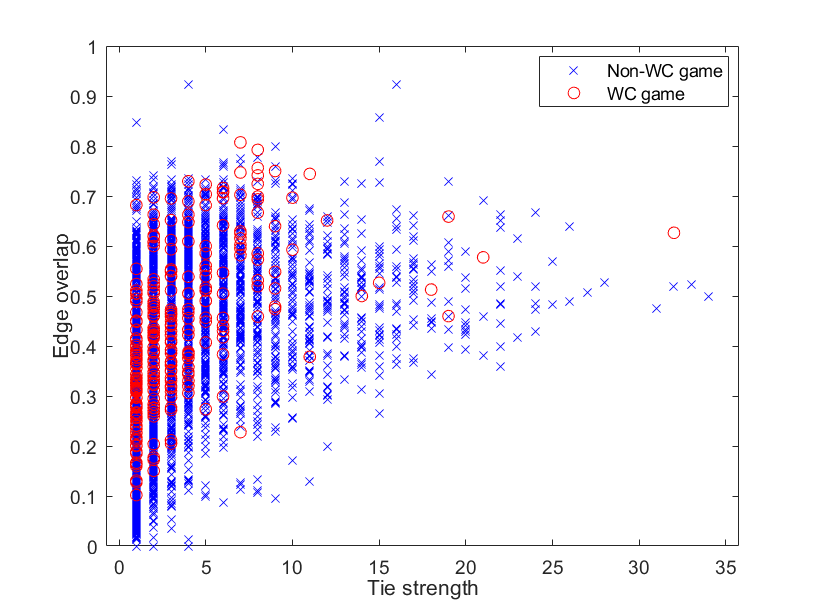}}
\subfigure[Tie strength vs. Edge overlap]{\label{fig4:b}\includegraphics[width = 90mm]{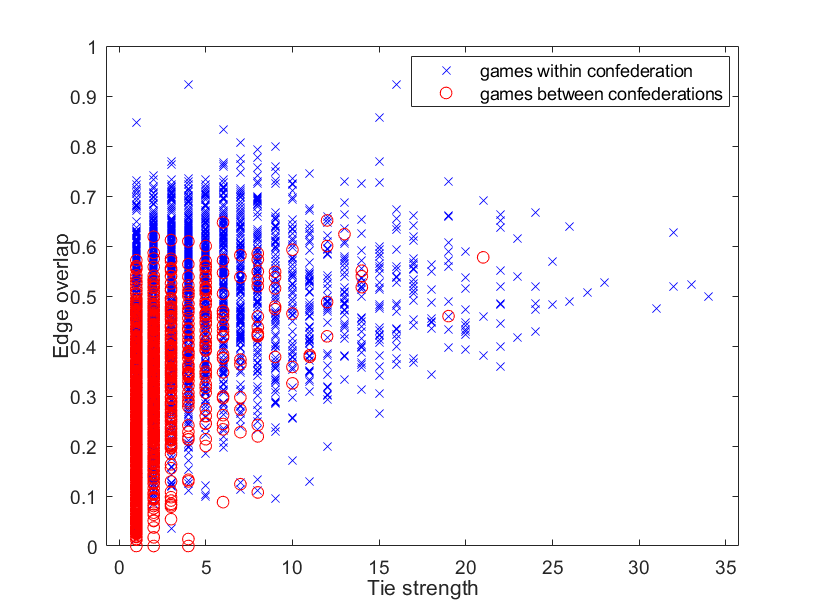}}
\caption{Tie strength vs. Edge overlap. Each point (circle/cross) corresponds to an edge with its tie strength as x-axis value and edge overlap as y-axis value. (a) World Cup (WC) games are marked as red circles and others are marked as blue crosses. Red circles gather in the lower left corner, indicating that most edges of World Cup games have low tie strength and low overlap; (b) Games between countries in different confederations are marked as red circles and games between countries in the same confederation are marked as blue crosses. Red circles gather in the lower left corner, indicating that most edges of games between confederations have low tie strength and low overlap}
\label{fig4:tieoverlap}
\end{figure}

In Fig.~\ref{fig4:tieoverlap}, we use the tie strength and edge overlap as the $(x,y)$ coordinates and plot all the edges between 1995 and 2015. In Fig.~\ref{fig4:a}, edges representing World Cup games are marked as red circles. Most red circles gather in the lower left corner, which indicates that most World Cup games have low tie strength and low overlap. As most World Cup games are played between countries from different continents, this observation again verifies the weak tie hypothesis in the football network where edges with low tie strength and low overlap are most likely between communities. Fig.~\ref{fig4:b} distinguishes games within and between confederations. Red circles that represent games between confederations are located near the origin, validating the existence of weak tie hypothesis in the football network. Fig.~\ref{fig4:b} also provides a visual correspondence to Table~\ref{t:fraction}, showing that most edges with low tie strength and low overlap exist between confederations. 

Fig.~\ref{fig3:edgeoverlap} and Fig.~\ref{fig4:tieoverlap} attest the weak tie hypothesis in the football network. As demonstrated in~\citep{granovetter1995getting}, most people know about their current jobs from an acquaintance instead of a friend. This fact reveals the role of weak ties in social cohesion and the vital importance of weak ties in message passing within social networks. Similarly, in the football network, edges with low tie strength are between countries with few games played between them. Edges with low overlap indicate that two countries do not share many common neighbors, which means there are few countries that these two countries have both played against. As a result, edges with either low tie strength or low overlap contain vital information about the structure of the football network, and serve as bridges between continents and confederations that contribute to integrate individual football societies into a complete, globalized football network. 

\begin{figure}[h!]
\centering     
\subfigure[Edge removal by tie strength]{\label{fig5:a}\includegraphics[width=60mm]{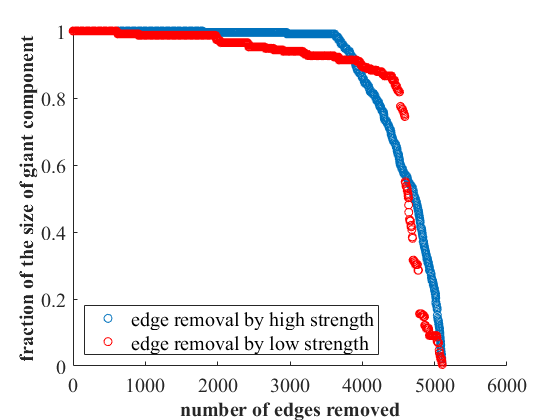}}
\subfigure[Edge removal by overlap]{\label{fig5:b}\includegraphics[width=60mm]{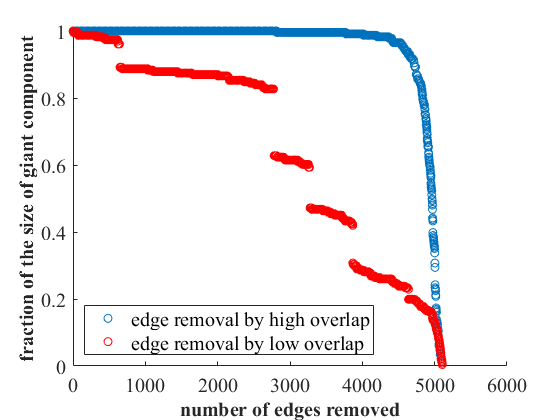}}
\caption{Change of the size of the giant component with edge removal. In both figures, the blue points correspond to removing the edge with the highest strength or overlap each time. The red points correspond to removing the edge with the lowest strength or overlap each time. The gaps between the red points indicate that removing edges with low strength or overlap will break down the graph and induce smaller communities}
\label{fig5:sizecomponent}
\end{figure}

To better reveal the community structures of the network using the concepts of tie strength and the edge overlap, edge removal was carried out by removing the strongest edge one by one. The relative size of the giant component (a connected subset of vertices whose size scales
extensively~\citep{newman2001scientific}) was computed to check the impact of the removal of each edge. Same procedure is repeated for removal of weakest edges as well. Fig.~\ref{fig5:a} and Fig.~\ref{fig5:b} show the size change of the giant component as edge removal is progressing. From the image, we can see that by progressively removing the edge with either the lowest tie strength or the lowest edge overlap , the size of the giant component shows discontinuity and gaps between points, indicating a sudden disintegration of the network. This means that removing edges with low strengths or overlap would lead to a breakdown of the original network and the emergence of multiple smaller communities. On the other hand, removing edges with high strengths and overlap gradually shrinks the whole network and does not in fact break the network apart.

In this section, the weak tie hypothesis has been validated, and the underlying community structures of the football network are revealed by edge removal. The next step is to formally detect these existing communities in the football network via community detection.

\subsection*{Community structure of static football network}\label{ss:staticcommunity}
\begin{figure}[h!]
\centering     
\subfigure[Adjacency matrix of the static football network]{\label{fig6:a}\includegraphics[width=90mm]{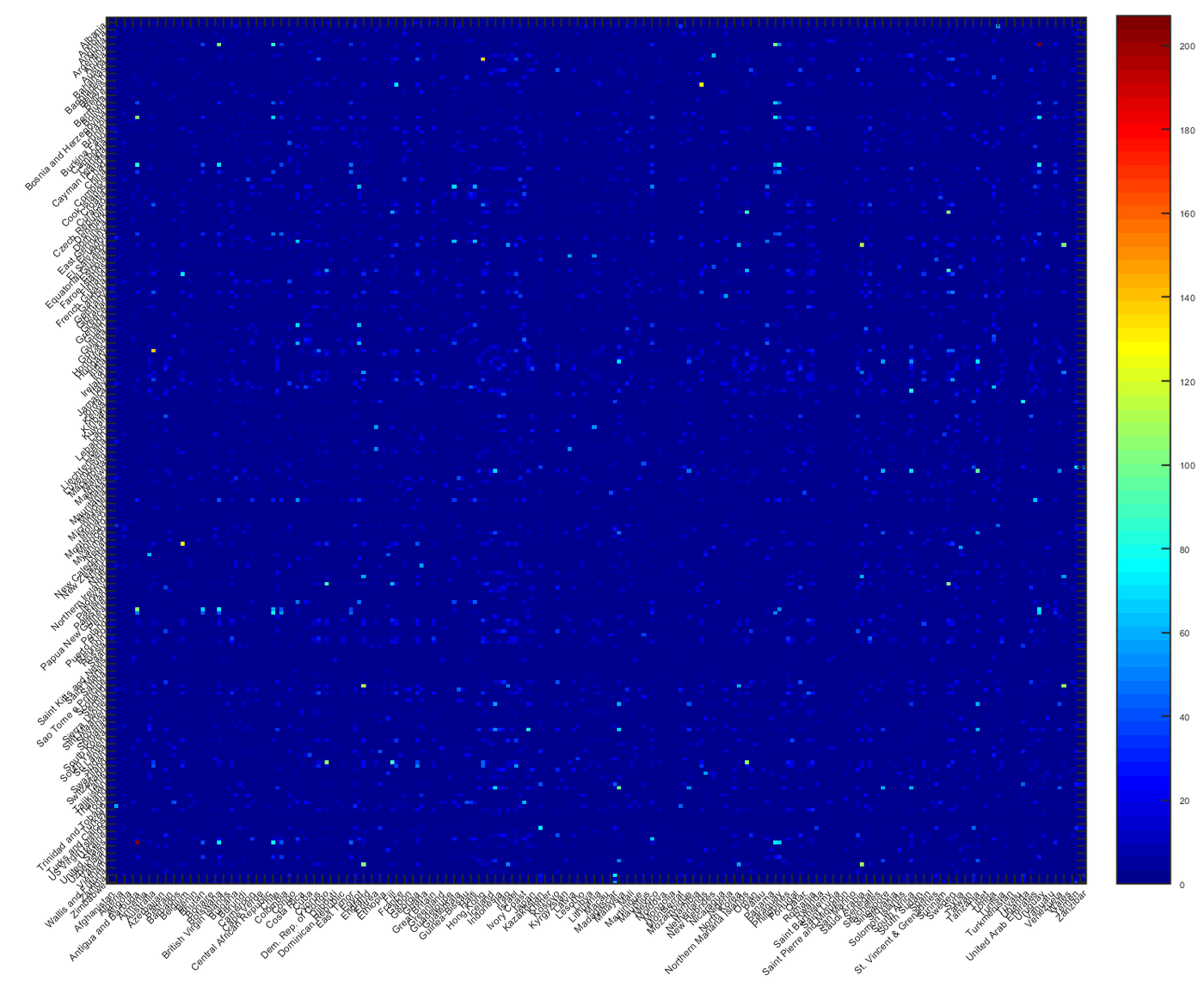}}
\subfigure[Adjacency matrix rearranged by node community assignment]{\label{fig6:b}\includegraphics[width=90mm]{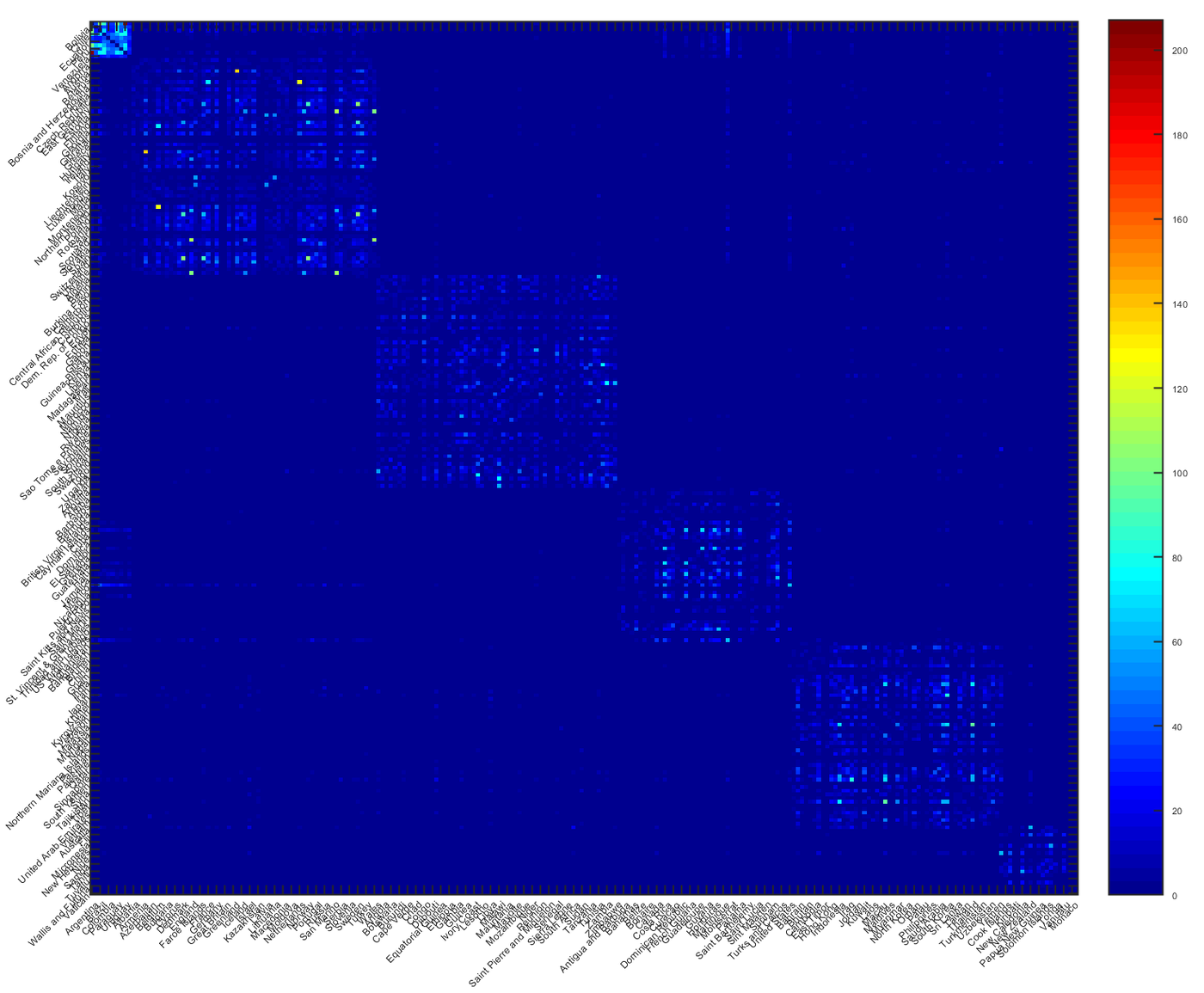}}
\caption{Adjacency matrix before and after community detection. (a) the original adjacency matrix of the representative network from 1872 to 2016; (b) the adjacency matrix in which the nodes are rearranged by their community assignments}
\label{fig6:adjmat}
\end{figure}

By using the complete data set from year 1872 to year 2016, we constructed a single representative graph of the football network. In this graph, 238 countries involved in football history are included as 238 nodes, and all the 39052 matches are represented by the edges. A $238 \times 238$ adjacency matrix shown in Fig.~\ref{fig6:a} was built for this network, whose entries are the number of matches played between two countries.

\begin{figure}[h!]
\centering
  \includegraphics[width=100mm, height = 60mm]{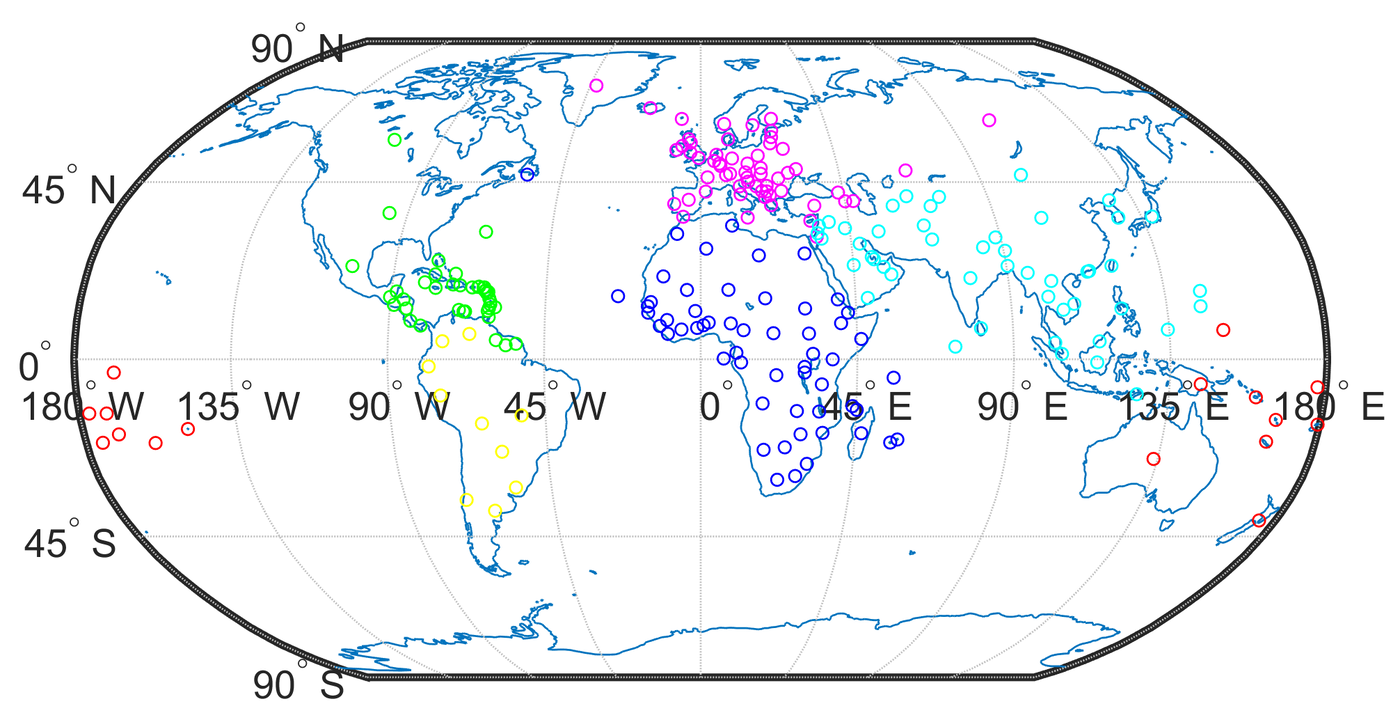} 
  \caption{Football community on the world map.
  Each color corresponds to one community. Based on the geographical locations of the circles on the map, we observe the following relationship between the confederations and the identified communities: purple: UEFA; blue: CAF; sky blue: AFC; red: OFC; green: CONCACAF; yellow: CONMEBOL}
  \label{fig7:worldmap}
  \end{figure}

Applying the Louvain community detection method on this network gives 6 detected communities. We rearrange the rows and columns in Fig.~\ref{fig6:a} based on the community assignment of each node to ensure that nodes of the same community are next to each other in the reformatted adjacency matrix. The new adjacency matrix exhibits a block-wise diagonal format as shown in Fig.~\ref{fig6:b}, with each block corresponding to one community. By plotting each community on the world map with different colors in Fig.~\ref{fig7:worldmap}, we see a clear correspondence between the detected communities and the football confederations presented in Table~\ref{t:FIFA}. 

Although Fig.~\ref{fig7:worldmap} shows a nice view of the structure of the football network, there exist a few exceptions. For example, Australia joined Asian Football Confederation in 2006 but is still included in the community corresponding to Oceania Football Confederation in Fig.~\ref{fig7:worldmap}. In addition, several countries such as Kazakhstan and Israel left AFC to join UEFA. Such transitions could not be revealed in a single network. Since more and more football matches take place in recent decades, it is reasonable to assume that football networks in earlier years are masked by the crowded networks of recent years where there are much more nodes and edges. The static football network is unable to reveal transitions and changes in the football society. In order to gain a dynamic view of the evolution of the football network, in the next section, we dissect the football history over time and dig into the dynamic properties of the football networks.

\subsection*{Dynamics of the football network}\label{ss:dynamics}
In this section, we focus on the dynamics of football networks with the aim to unveil the evolution and globalization of football society.

\subsubsection*{Descriptive statistics of dynamic networks}\label{sss:graphmeasures}
To obtain the temporal dynamics of the football networks, we extract all the football match records from year 1901 to year 2010, and group them into 11 decades to generate one representative football network for each decade. We compute the number of games played either within each football confederation or between confederations per decade. The intention is to distill appealing information, such as which confederations dominate the football world, which confederations experience sudden prosperity or stasis, etc. It would also be interesting to correspond the observations with specific historical events. For example, we would expect to see a significant decrease in the amount of games played in UEFA due to the war, and a sudden spike in the 1950s for CAF as Africa officially enters the football world.

\begin{figure}[h!]
\centering     
\subfigure[Number of games within confederations]{\label{fig8:a}\includegraphics[width=90mm]{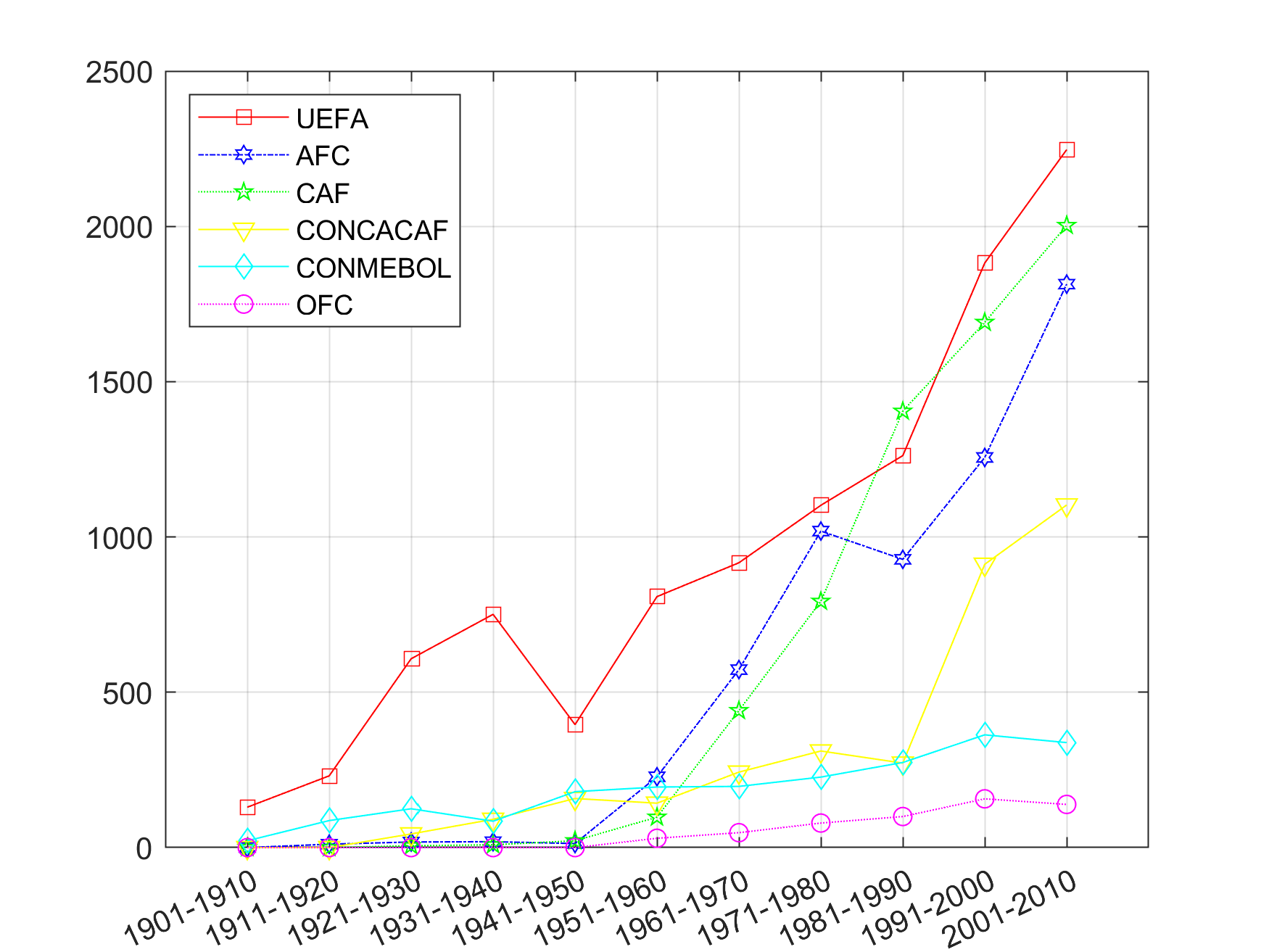}}
\subfigure[Number of games between confederations]{\label{fig8:b}\includegraphics[width=90mm]{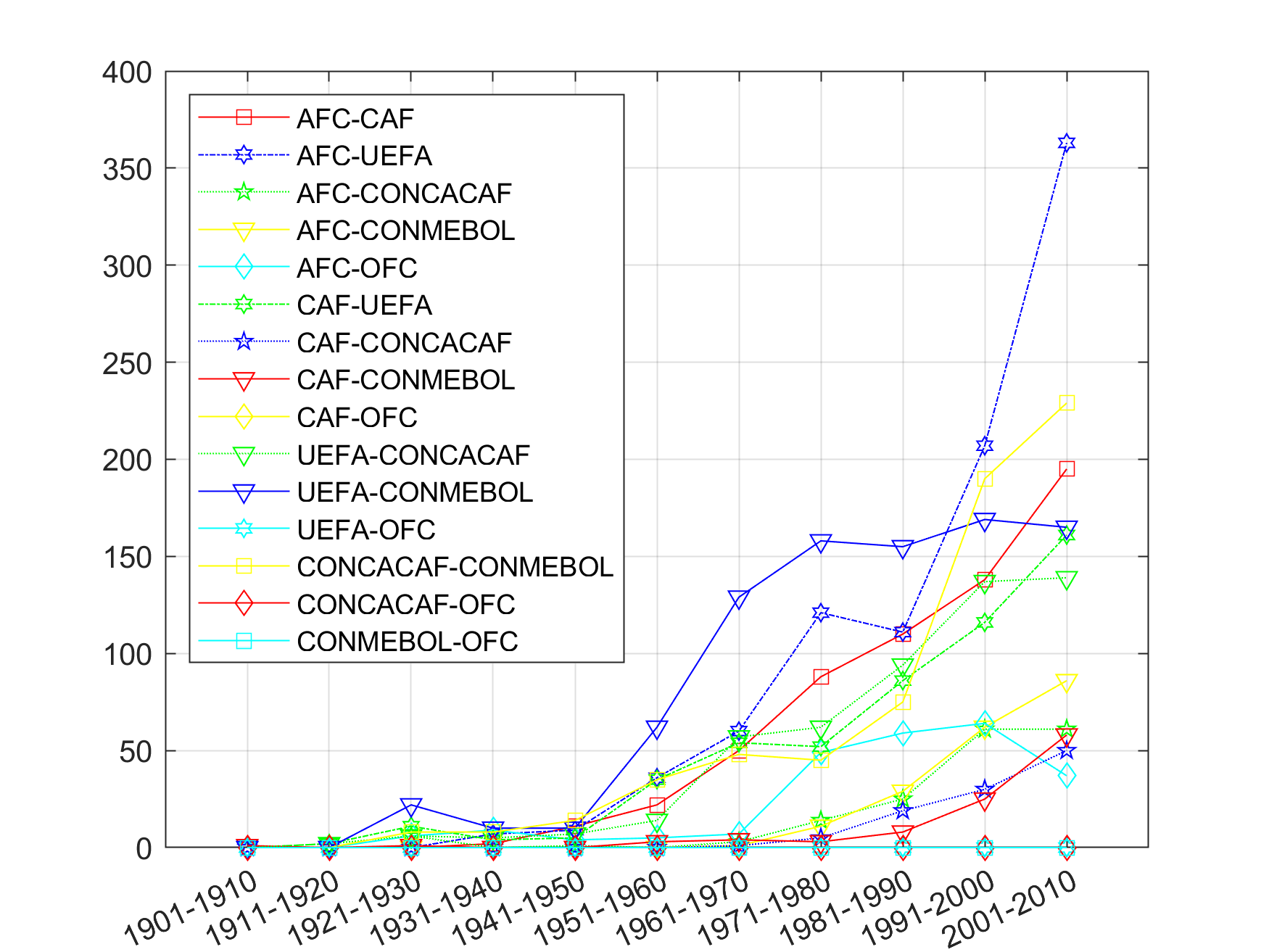}}
\caption{Number of edges within/between confederations per decade. (a) total number of football games played within each confederation for each decade; (b) total number of football games played between any two confederations for each decade}
\label{fig8:edgecount}
\end{figure}

Different football confederations have different development paths. Some entered the football world early while others had a late start. Fig.~\ref{fig8:edgecount} shows the number of football games played within and between confederations. In Fig.~\ref{fig8:a}, it is clear that the dominant confederation is UEFA shown as the red line with the most games played. It suffered a severe drop down in the number of games in the decade of 1941-1950 due to the second World War. CONMEBOL shown as the sky blue diamond, as the first established confederation, does not experience much interruption and shows a steady growth. AFC and CAF do not have many football games until the 1950s, the decade in which both confederations were established. The first Asian Cup and the first African Cup were also held in that decade.

In addition, it is worth noticing that the 1990s witnesses great increases in number of games played in multiple confederations. Such increase makes sense if we look into football history for reference. The 1998 World Cup grew from 24 teams to 32 teams and allowed more teams from Africa, Asian and North America to participate. This change could significantly increase the eagerness of countries in these areas to join football and also enhance the competition. More friendly games and qualification games would be played within confederations. 

Fig.~\ref{fig8:b} shows the interaction between confederations. The communication between confederations is basically growing as more games are played in recent decades. Exception such as the number of games between AFC and OFC, shown as the sky blue line with diamonds, can be explained as Australia left OFC and joined AFC in 2006. Consequently, the original between-confederation edges between Asian countries and Australia now belong to the within-confederation edges of AFC. The line with the blue star shows a significant increase of football games between AFC and UEFA starting from the 1990s. This change is strongly related to the ambition of Asian countries in developing their football and the growing economy of Asia where money are spent to invite European teams for friendly games.

Other graph measures can also be useful to capture the dynamics of networks. In this work, we explore the measure of global efficiency~\citep{latora2001efficient} which is the average of inverse shortest path length. The average efficiency of a network \textit{G} is defined as:
\[
 E(G) = \frac{1}{n(n-1)}\displaystyle\sum_{i<j \in G}\frac{1}{d_{ij}},
\]
where $n$ denotes the total number of nodes and $d_{ij}$ denotes the length of the shortest path between node $i$ and node $j$. The global efficiency is defined as:
\[
 E_{global}(G) = \frac{E(G)}{E(G^{ideal})}
\]
where $G^{ideal}$ is the graph with $n$ nodes and all possible edges are present. Global efficiency serves as a quantitative measure of the average distance it takes for a node to reach another node. Networks with high global efficiency should have more edges thus connections between nodes are efficient.

\begin{figure}[h!]
\centering
\includegraphics[width=90mm, height = 60mm]{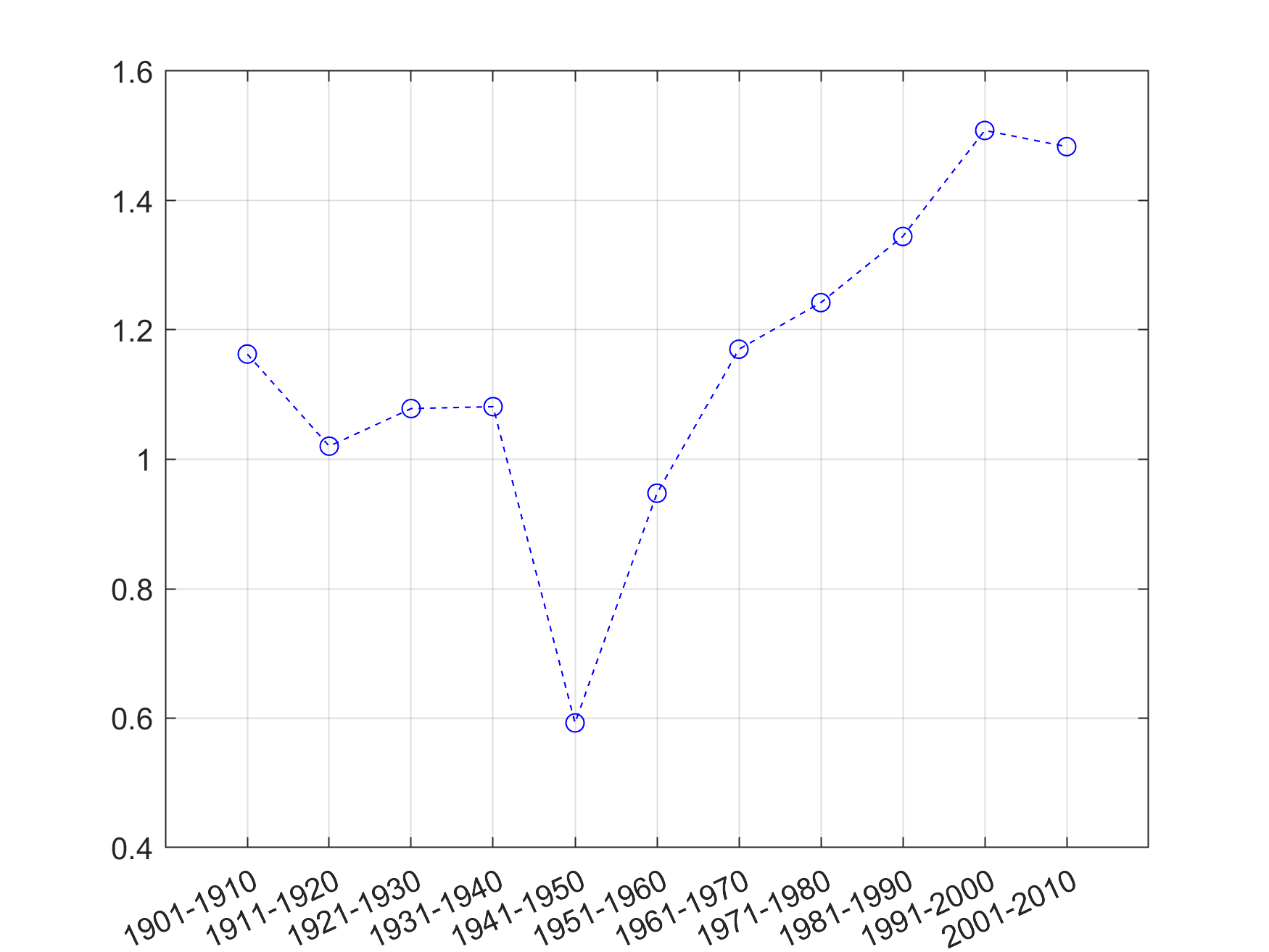} 
\caption{Global efficiency of the football networks per decade. An ascending trend is witnessed, with the exceptions due to the world wars}
\label{fig9:globaleff}
\end{figure}

Fig.~\ref{fig9:globaleff} shows the computed global efficiency for the football networks constructed for each decade. An obvious valley appears in the 1940s when the WWII broke out. This finding matches the results in Fig.~\ref{fig8:edgecount}. During the wartime, there were fewer football games thus most connections between the nodes were cut off, and the global efficiency of the network is severely compromised.

So far, we have looked at several descriptive statistics of the dynamic in the football networks. In Fig.~\ref{fig7:worldmap}, we observe the correspondence between football communities and real-world football confederations. However, football networks do not always appear in such form as different confederation were established at different times in history. In addition, it is practical to assume that communities in early decades are more localized in certain regions, while in recent decades communities are much more spread out and nodes in the same community may have huge geographical distance in between. Thus, it is reasonable to assume that football networks maintain different community structures in different decades.

\subsubsection*{Community structure of dynamic networks}\label{sss:dynamicstructure}
\begin{figure}[h!]
\centering     
\subfigure[1901-1910]{\label{fig10:a}\includegraphics[width=60mm]{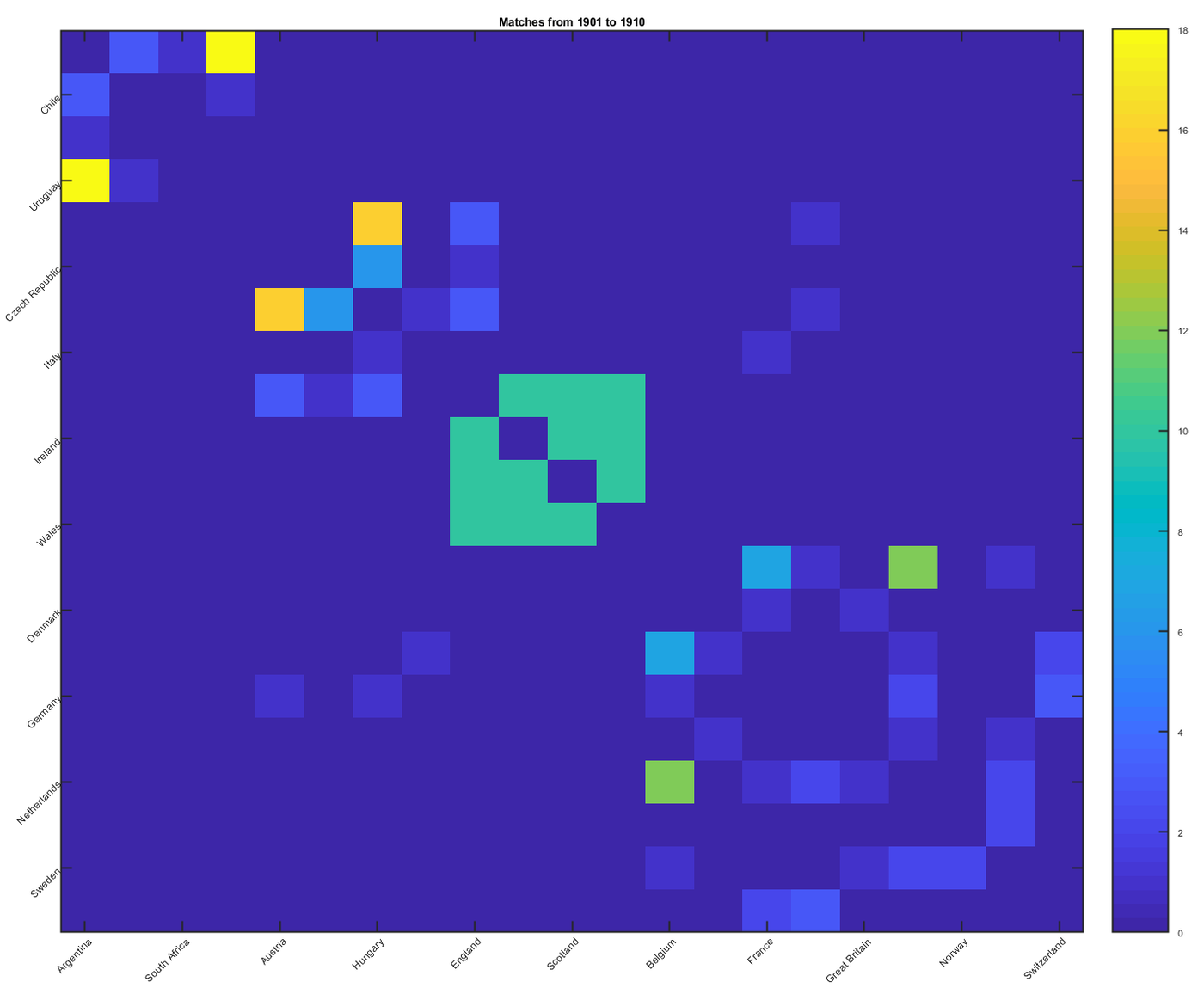}}
\subfigure[1901-1910]{\label{fig10:b}\includegraphics[width=60mm]{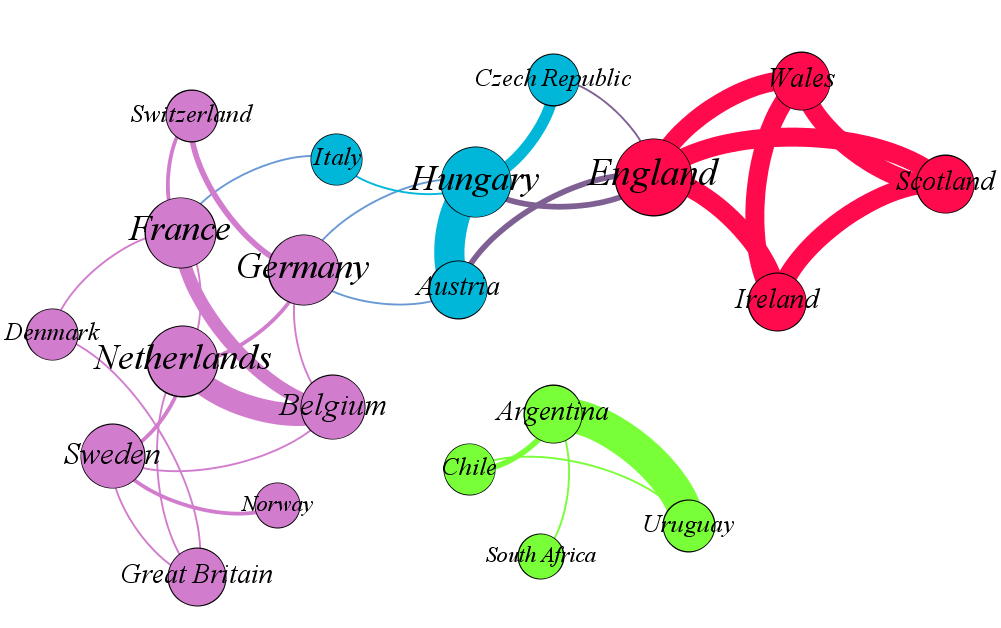}}
\subfigure[1931-1940]{\label{fig10:c}\includegraphics[width=60mm]{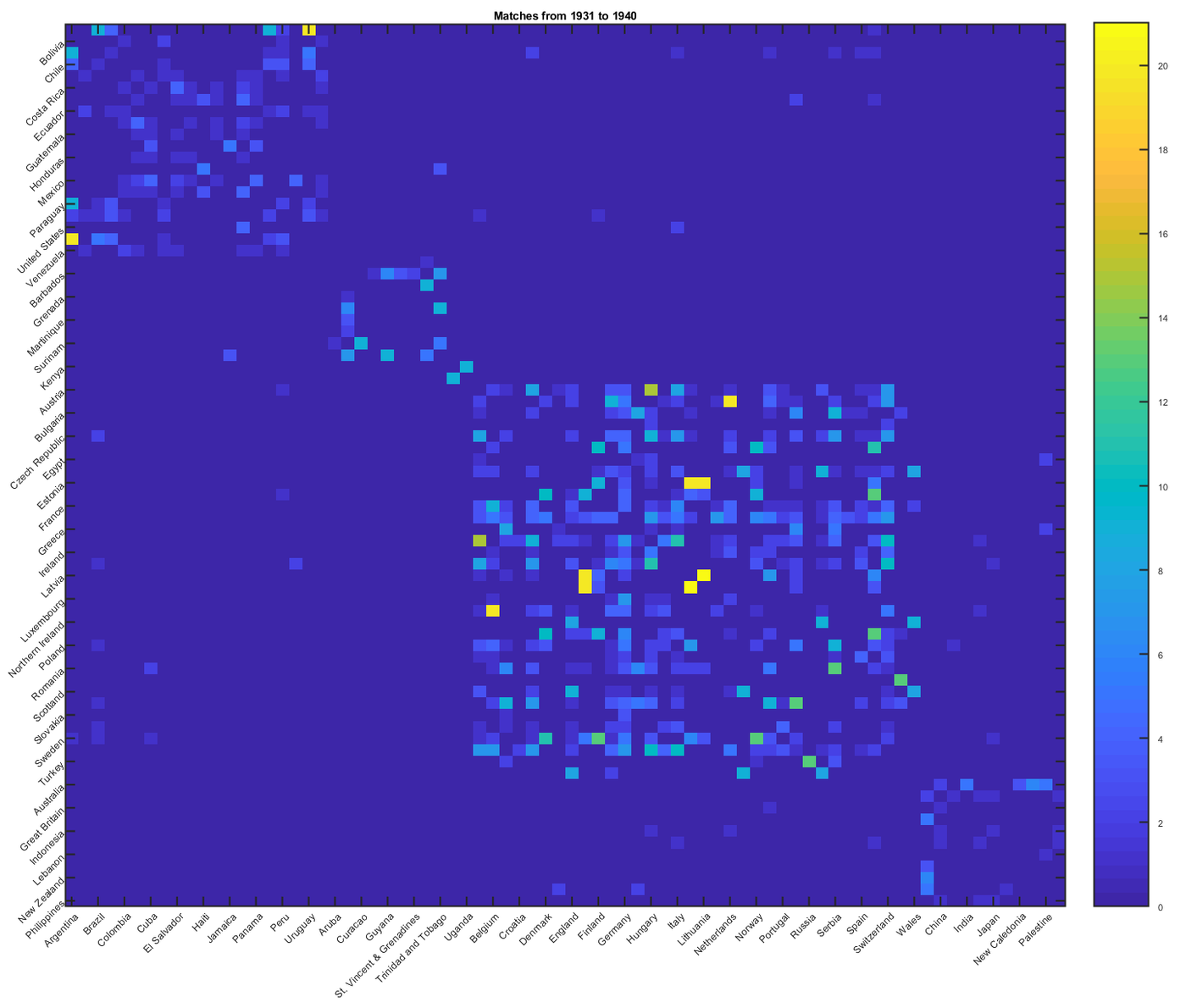}}
\subfigure[1931-1940]{\label{fig10:d}\includegraphics[width=60mm]{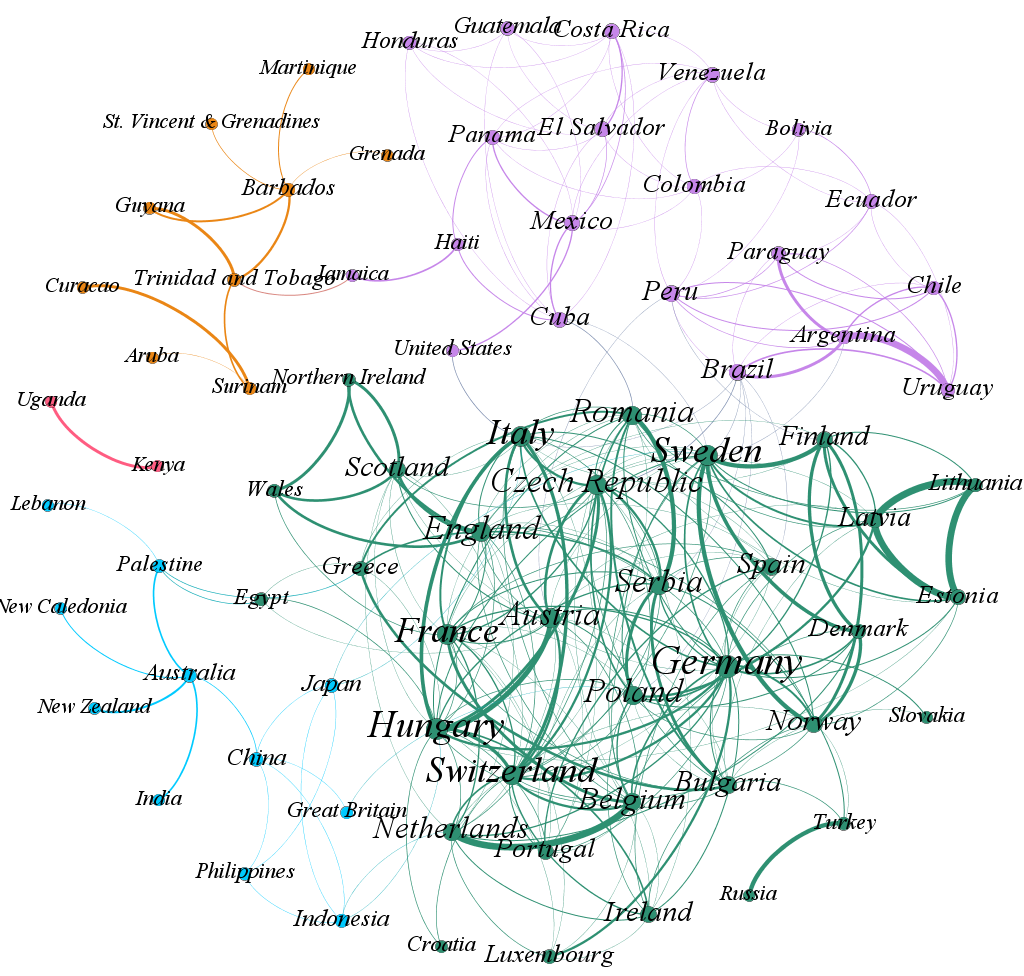}}
\caption{Adjacency matrices and visualization for decade 1901-1910 and 1931-1940. Left: adjacency matrices. Right: network visualization. In (b) and (d), each color stands for one community. In early decades, boundaries between communities in the adjacency matrices as shown in (a) and (c) were not clear, and the correspondence between community assignment and confederation as shown in (b) and (d) was poor}
\label{fig10:communityevolution}
\end{figure}

\begin{figure}[h!]
\centering     
\subfigure[1971-1980]{\label{fig11:a}\includegraphics[width=60mm]{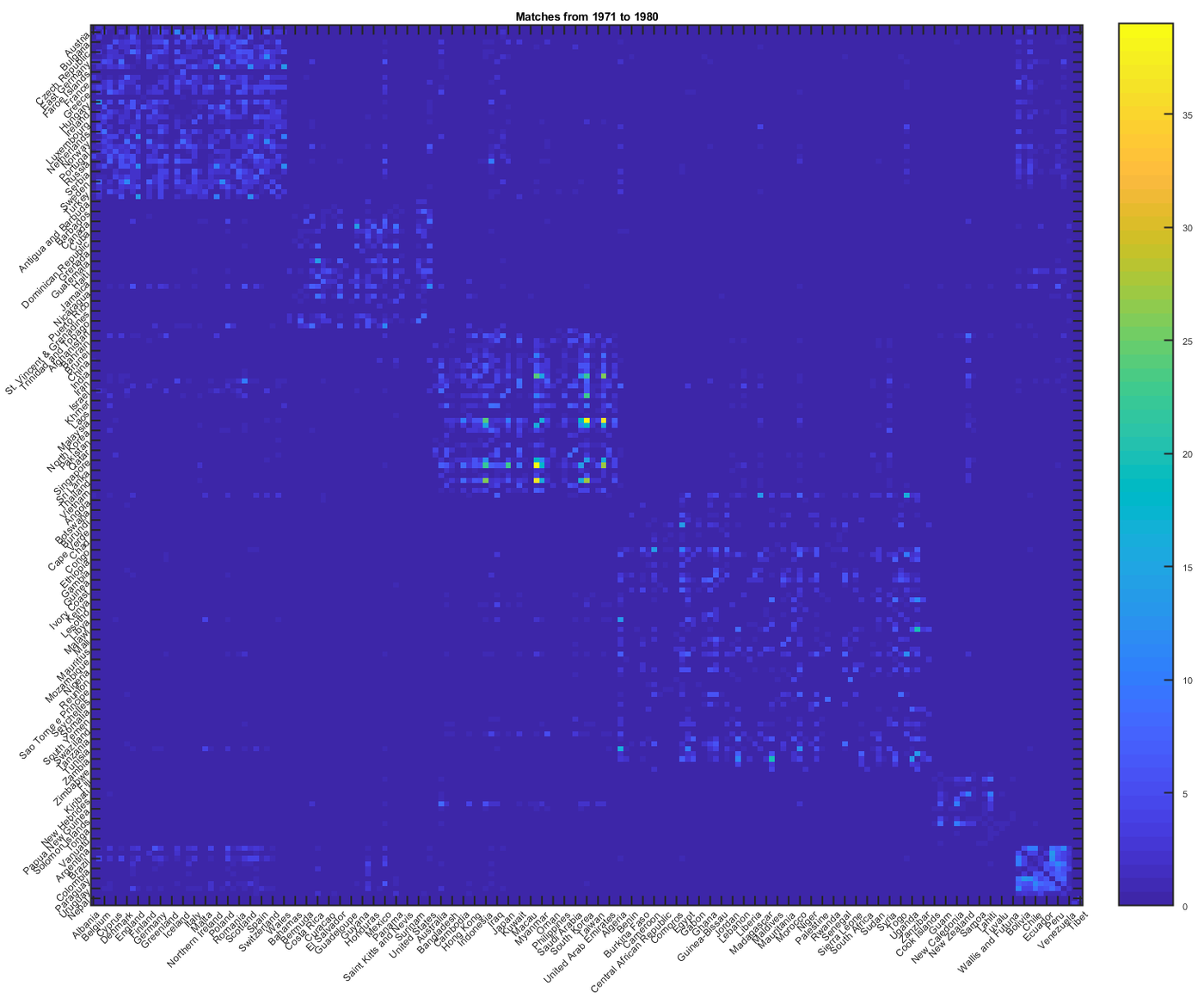}}
\subfigure[1971-1980]{\label{fig11:b}\includegraphics[width=60mm]{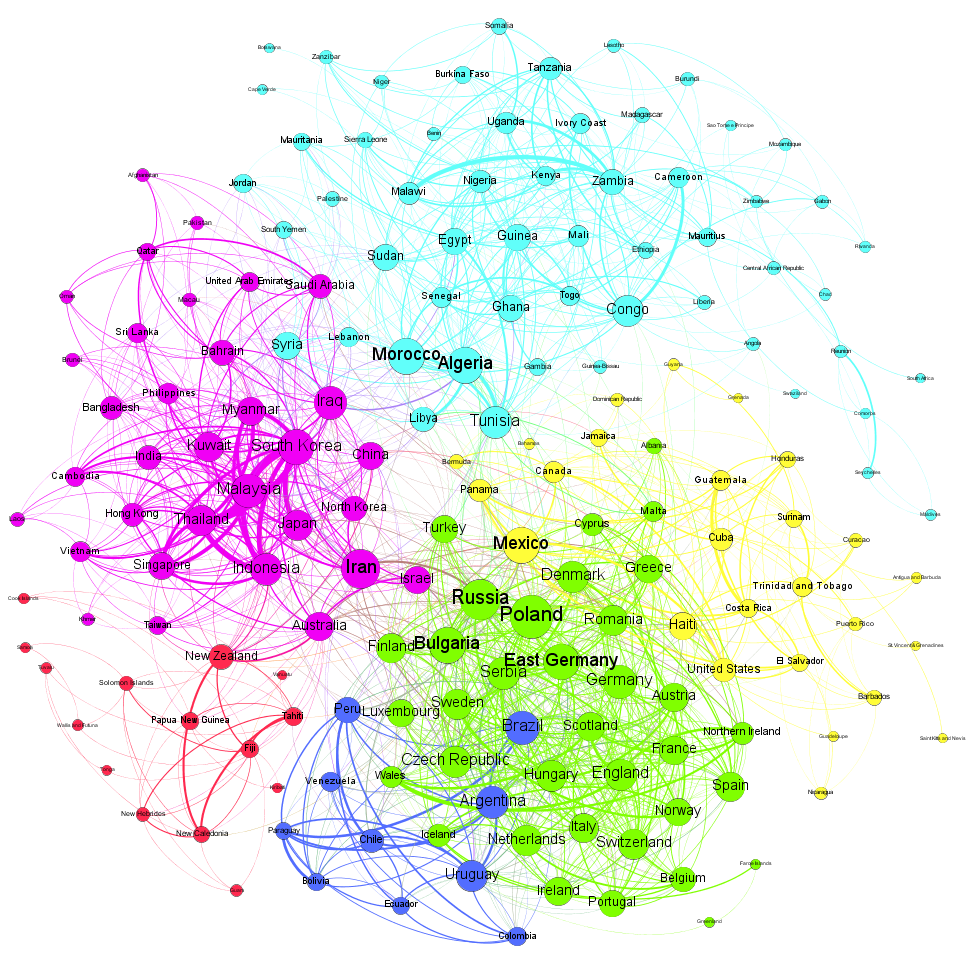}}
\subfigure[2001-2010]{\label{fig11:c}\includegraphics[width=60mm]{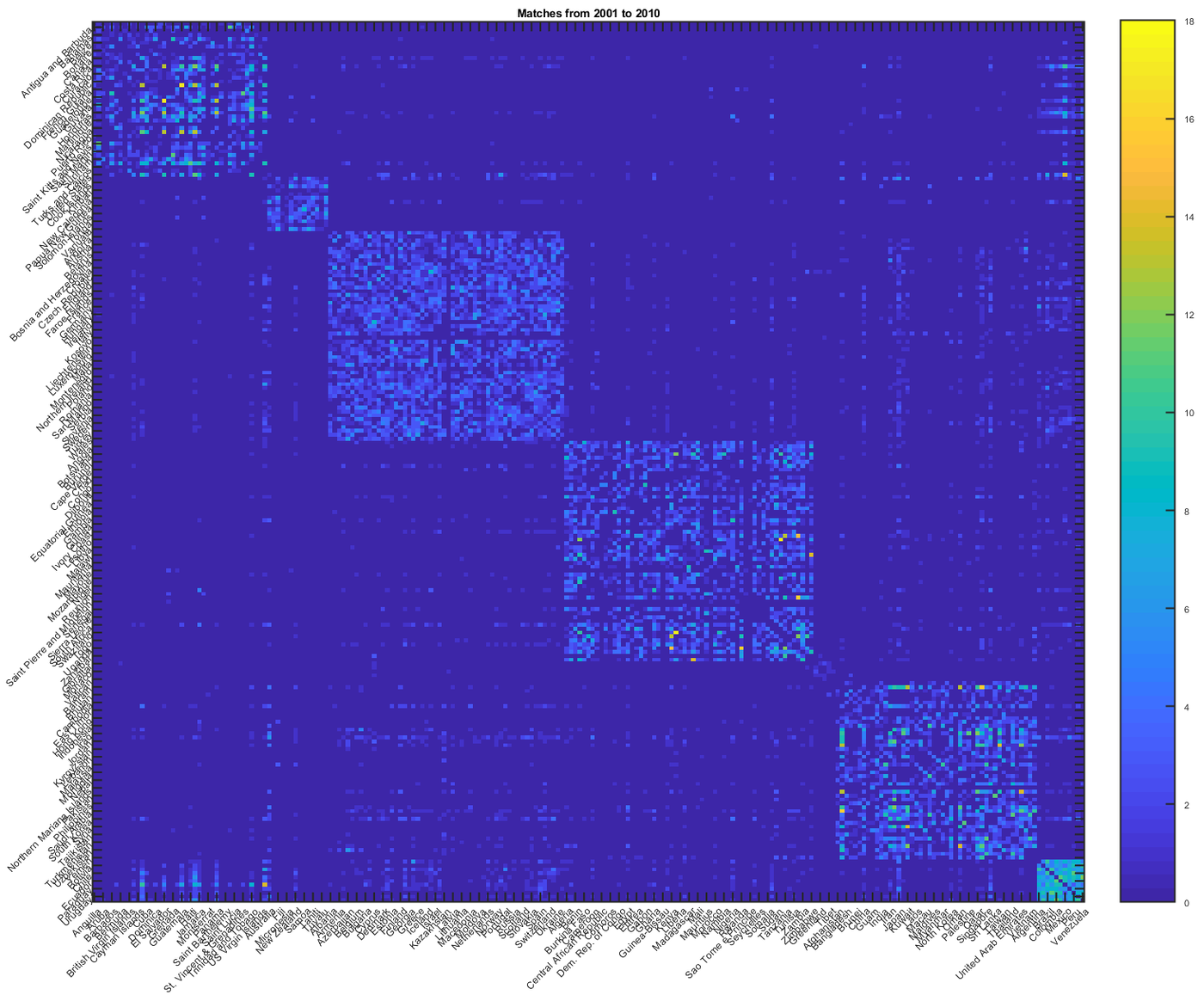}}
\subfigure[2001-2010]{\label{fig11:d}\includegraphics[width=60mm]{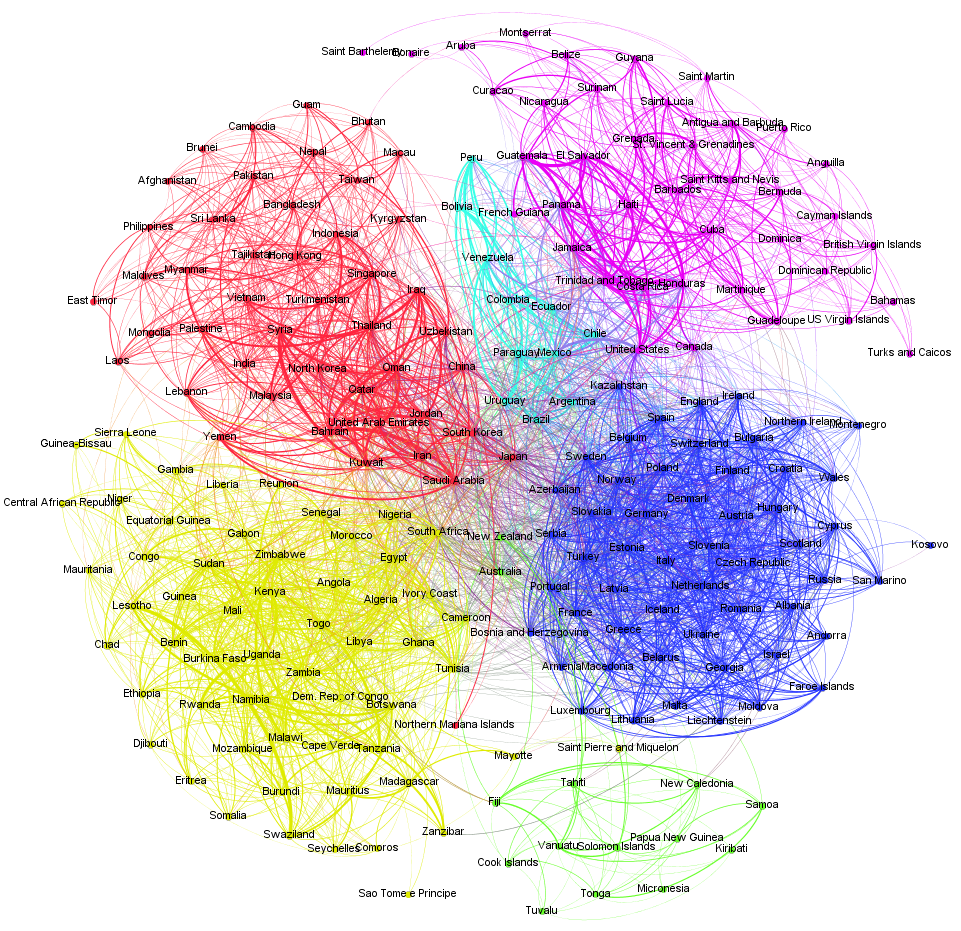}}
\caption{Adjacency matrices and visualization for decade 1971-1980 and 2001-2010. Left: adjacency matrices. Right: network visualization. Networks get more crowded and complex with more nodes and edges in recent decades, and the community structure is more similar with the modern football landscape represented by the six confederations}
\label{fig11:communityevolution2}
\end{figure}

To reveal the dynamics of the community structure of football networks, we applied the Louvain community detection algorithm on the 11 networks for the 11 decades from year 1901 to year 2010. Fig.~\ref{fig10:communityevolution} and Fig.~\ref{fig11:communityevolution2} give the adjacency matrices and visualization of the networks for 4 example decades. Both figures show a clear trend of the community evolution of the football network. Starting from early decades, the communities are quite local, and the correspondence between the identified communities and confederations was not clear. As more countries join the football society, the communities start to grow with more nodes. The boundaries between communities also become more apparent as shown in Fig.~\ref{fig10:c} and Fig.~\ref{fig11:a}. In these decades, the communities are slowly transforming their structures, sharing more and more similarity with the actual football confederations. Fig.~\ref{fig11:c} and~\ref{fig11:d} show the football network of the last decade, in which the community assignment for each node is basically the same with the actual confederation affiliation of each country. From these figures, a clear evolution path of the football network is unveiled, and the temporal features beneath such evolving community structures definitely worth more in-depth research.

\subsubsection*{Temporal states extraction}\label{sss:temporalstates}
In the previous section, we brought up the assumption that different stages exist during the evolution of football. It would be helpful to identify individual states in football history which represent different evolution stages. As stated in multiple literatures~\citep{zager2008graph,papadimitriou2010web,koutra2013deltacon}, graphs belonging to the same state or same cluster shall exhibit high similarity. As a result, a reasonable way to find temporal states in football history is to first calculate the similarity between football networks per year, and gather the graphs with high similarities into one group.

Multiple literatures~\citep{bunke2000graph,macindoe2010graph,bunke2007graph,wilson2008study} have discussed the calculation of graph similarity. However, each method only consider one aspect of the graph features and may lose other useful information. In~\citep{li2011graph}, the authors included 20 features regarding graph properties to generate a feature vector per graph. After normalizing the feature vectors, they fed them into a SVM for graph classification. The mean of the node degrees and the mean of node clustering coefficients are combined with other global measures such as the global efficiency in the final feature vectors. This procedure ignores the node correspondence between graphs, and may cause information loss regarding the different degree distribution between graphs. Also, combining all the measures into a singe vector makes it vague to determine the contribution from each measure. In order to preserve the node correspondence and take advantage of the graph-level measures, we bring up the following framework to combine different categories of graph measures for the computation of graph similarities. We continue to use the match records from year 1901 to year 2010, and construct 110 football networks in total, one for each year. The graph measures are categorized as the following 3 types.

(1) Node-level: degree, clustering coefficient~\citep{watts1998collective}, closeness~\citep{freeman1978centrality}, local efficiency~\citep{latora2001efficient}.  
\begin{itemize}
\item For each graph with $N$ nodes, compute each of the above measures and generate a $N \times 1$ vector for each node-level measure;
\item Concatenate the vectors for all the 110 networks into a $110 \times N$ matrix;
\item Calculate the correlation coefficient between matrix rows;
\item Transform the correlation coefficient $c_{ij}$ into similarity measure to constrain the value into [0,1];
\[
 s_{ij} = \frac{c_{ij}+1}{2},i,j = 1,...,110
\]
\item Generate a $110 \times 110$ similarity matrix for each node-level measure.
\end{itemize}

(2) Graph-level: number of nodes, number of edges, average path length, global efficiency~\citep{latora2001efficient}, diameter, radius, graph energy~\citep{gutman1978energy}, link density~\citep{black2008sparse} and transitivity~\citep{newman2003ego}.
\begin{itemize}
\item Calculate the above measures and construct a $9 \times 1$ feature vector per graph;
\item For all the 110 graphs, construct a $110 \times 9$ feature matrix;
\item Normalize the columns with z-normalization;
\item Generate a $110 \times 110$ similarity matrix based on the similarity (defined above) between rows.
\end{itemize}

(3) Structure-level: vertex-edge-overlap (VEO)~\citep{papadimitriou2010web}. The vertex-edge-overlap similarity is defined as
\[
 sim_{VEO}(G_1,G_2)=2\times\frac{E_1\cap E_2+V_1\cap V_2}{E_1+E_2+V_1+V_2}
\]
where $G_1 = (V_1,E_1),G_2 = (V_2,E_2)$. The vertex-edge-overlap measures the structural similarity between graphs. The VEO is computed between every pair of graphs of the 110 graphs, and a $110 \times 110$ similarity matrix is generated.

\begin{figure}[h!]
\centering
\includegraphics[width=120mm, height = 160mm]{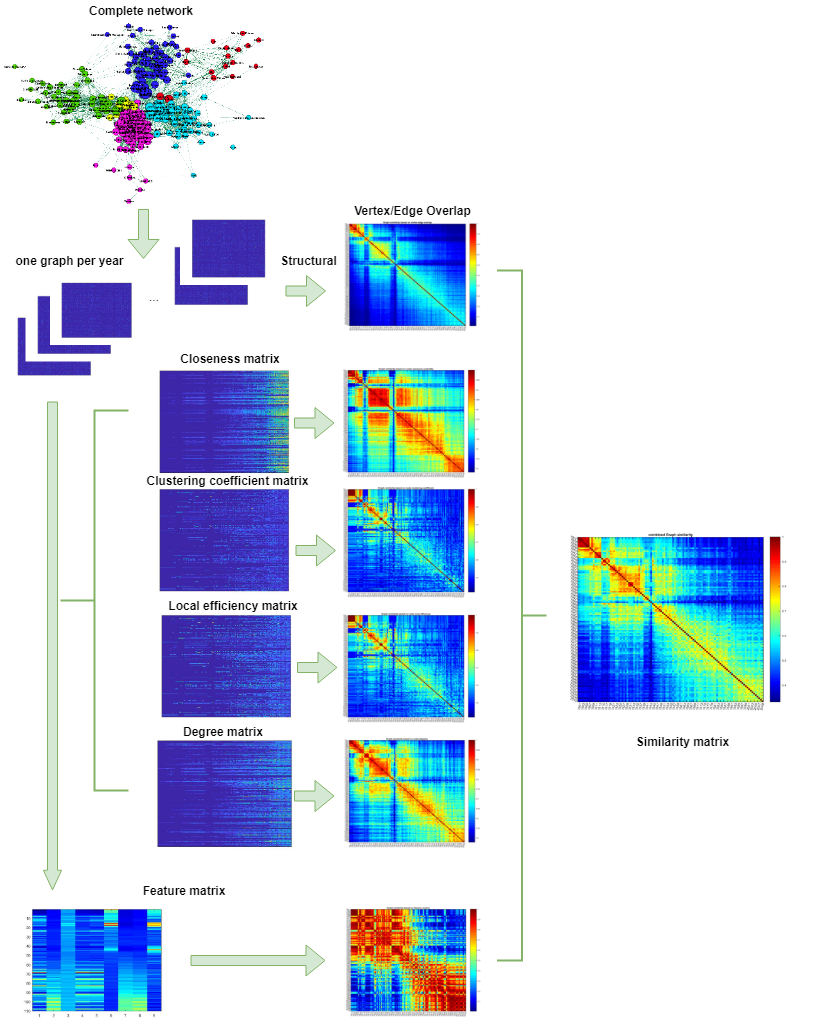} 
\caption{Pipeline for identifying temporal states in football history. 110 networks are generated for each year from 1901 to 2010. Node-level, graph-level, structure-level similarity matrices are computed and the final similarity matrix is given on the right}
\label{fig12:pipeline}
\end{figure}

After the three similarity matrices are obtained, we further calculate the average of them as the final similarity matrix. Fig.~\ref{fig12:pipeline} presents the exact pipeline for the above procedure to identify temporal states in football history. This framework takes into consideration the node-level graph measures to preserve the node correspondence between graphs, the global graph measures to consider overall graph properties, and the structural properties (nodes and edges) to quantify the topological similarity between graphs. In Fig.~\ref{fig12:pipeline}, a clear dissection of the original complete network in the top left corner is presented. Three levels of similarity calculation are carried out and the results shall be discussed soon in later sections.

\begin{figure}[h!]
\centering     
\subfigure[degree]{\label{fig13:a}\includegraphics[width=60mm]{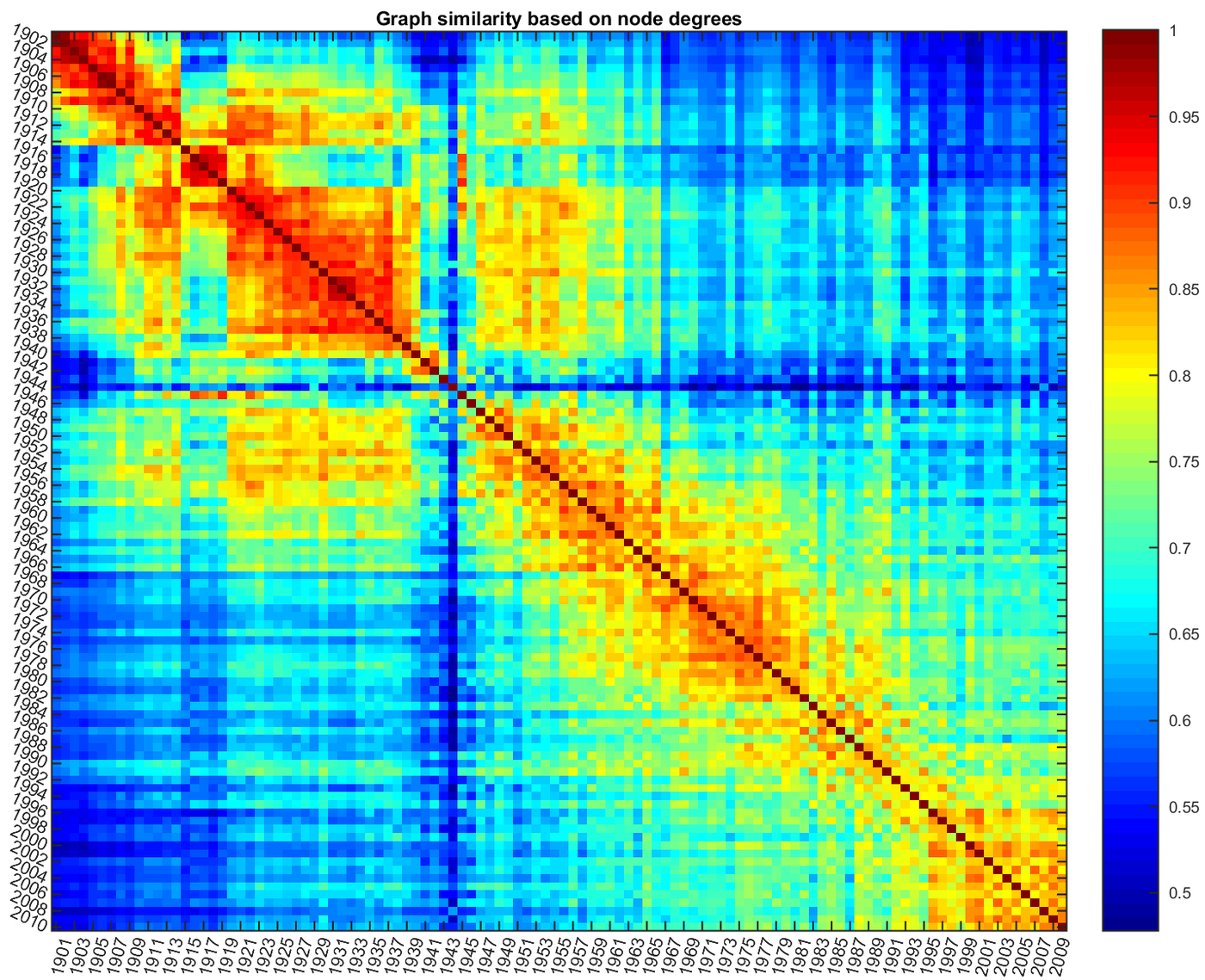}}
\subfigure[closeness]{\label{fig13:b}\includegraphics[width=60mm]{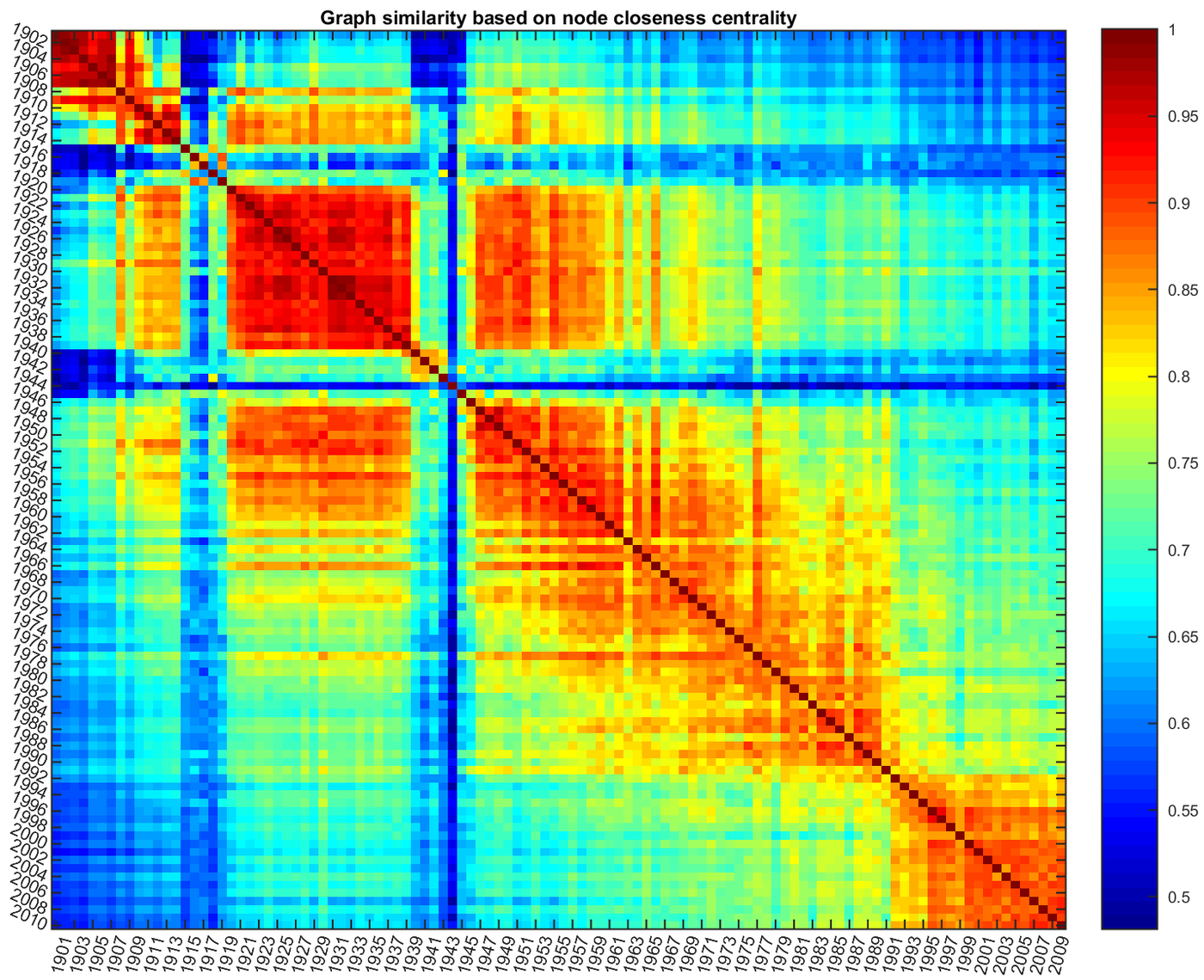}}
\subfigure[clustering coefficient]{\label{fig13:c}\includegraphics[width=60mm]{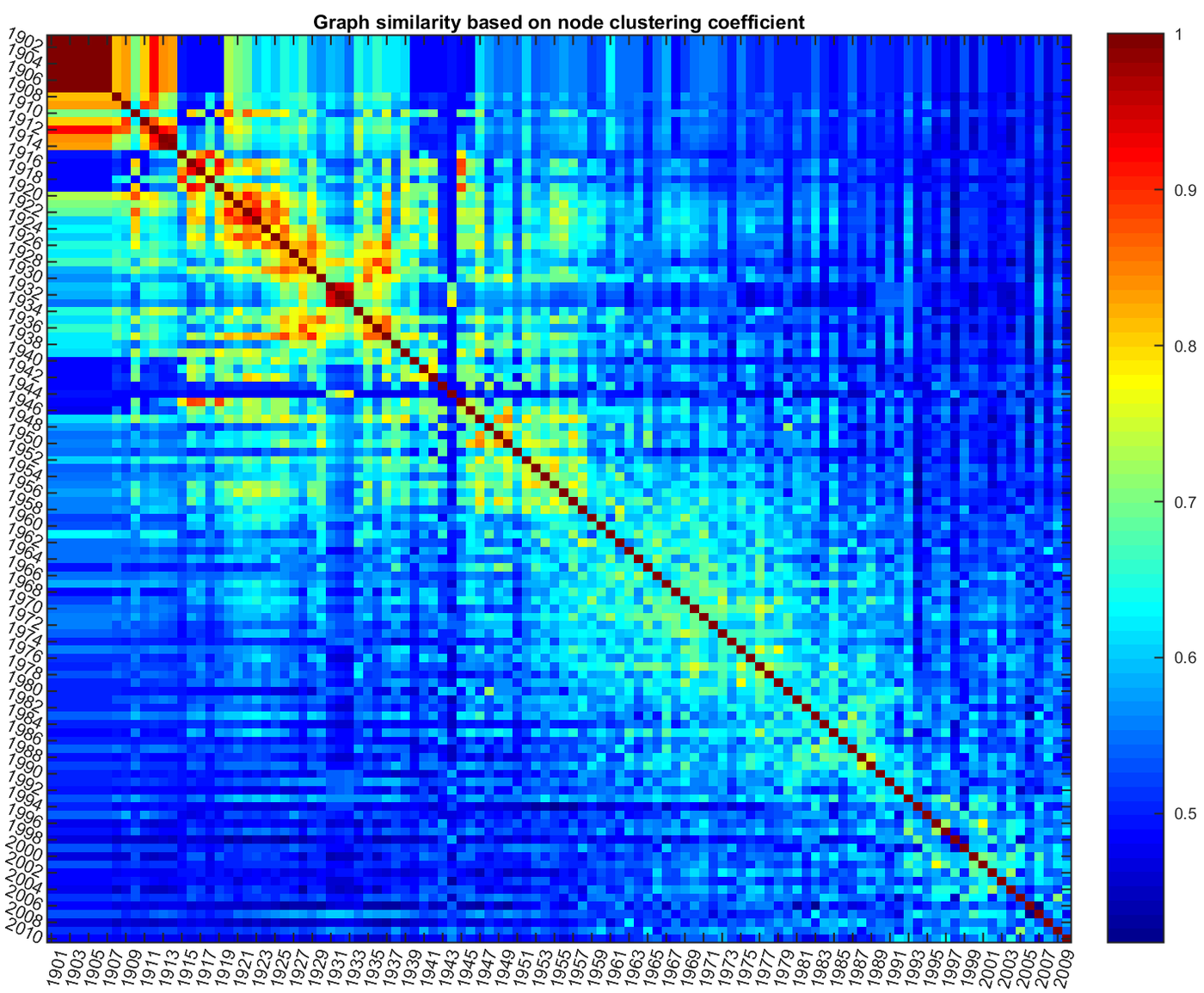}}
\subfigure[local efficiency]{\label{fig13:d}\includegraphics[width=60mm]{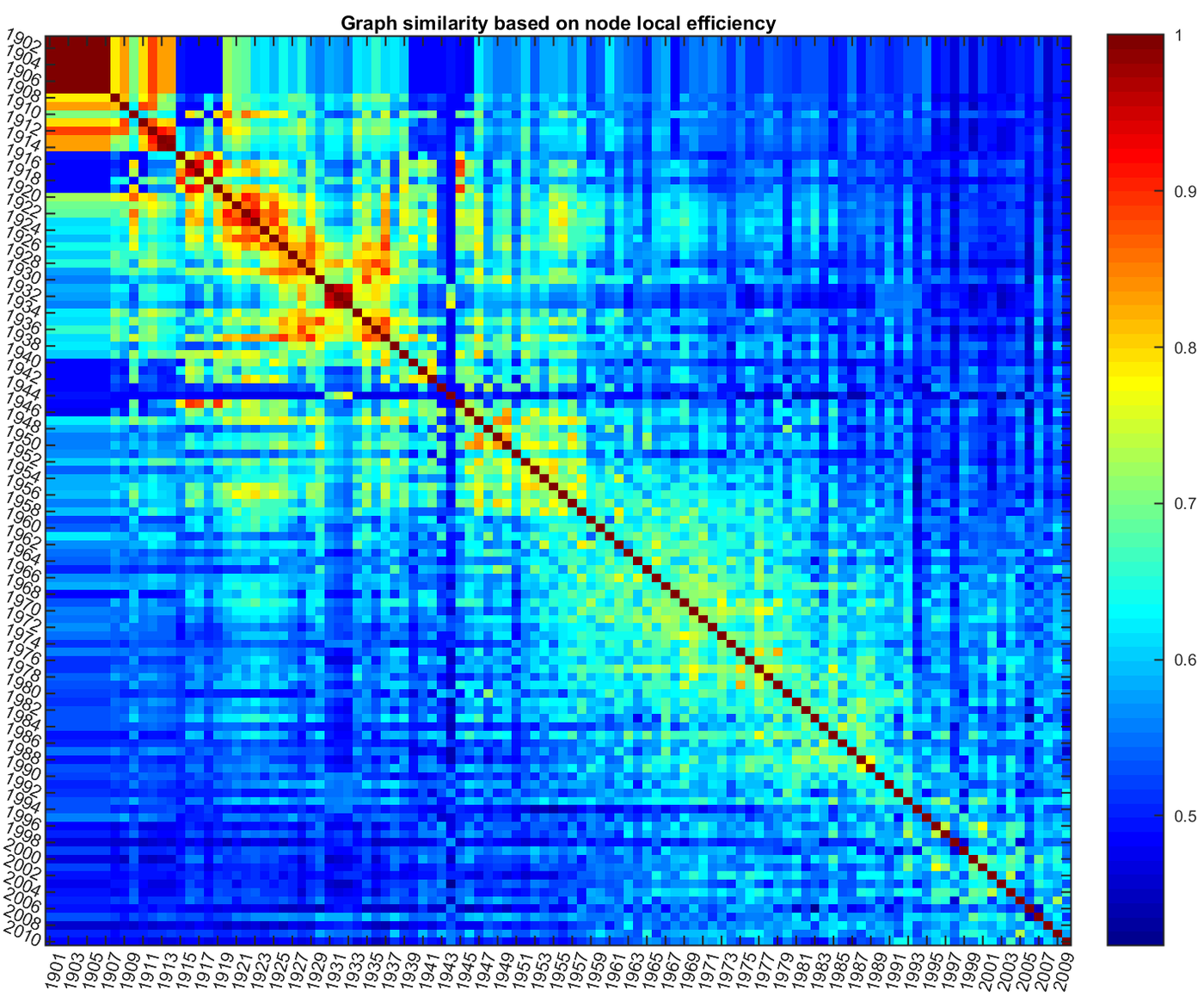}}
\caption{Graph similarity matrices based on node-level graph measures. Similarity matrix based on (a) node degree (b) node closeness (c) node clustering coefficient (d) node local efficiency}
\label{fig13:nodelevel}
\end{figure}

\begin{figure}[h!]
\centering     
\subfigure[global measures]{\label{fig14:a}\includegraphics[width=60mm]{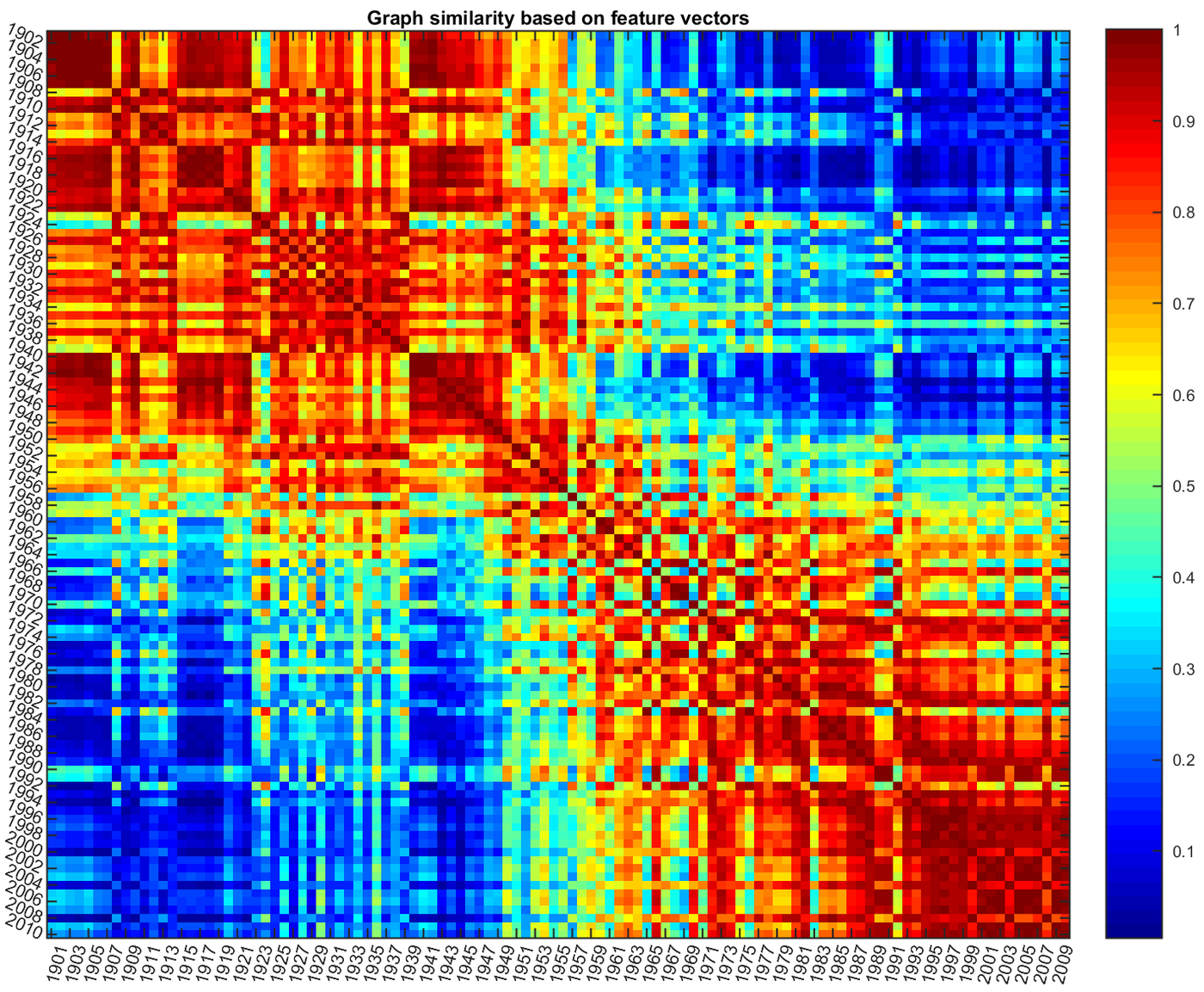}}
\subfigure[vertex-edge-overlap]{\label{fig14:b}\includegraphics[width=60mm]{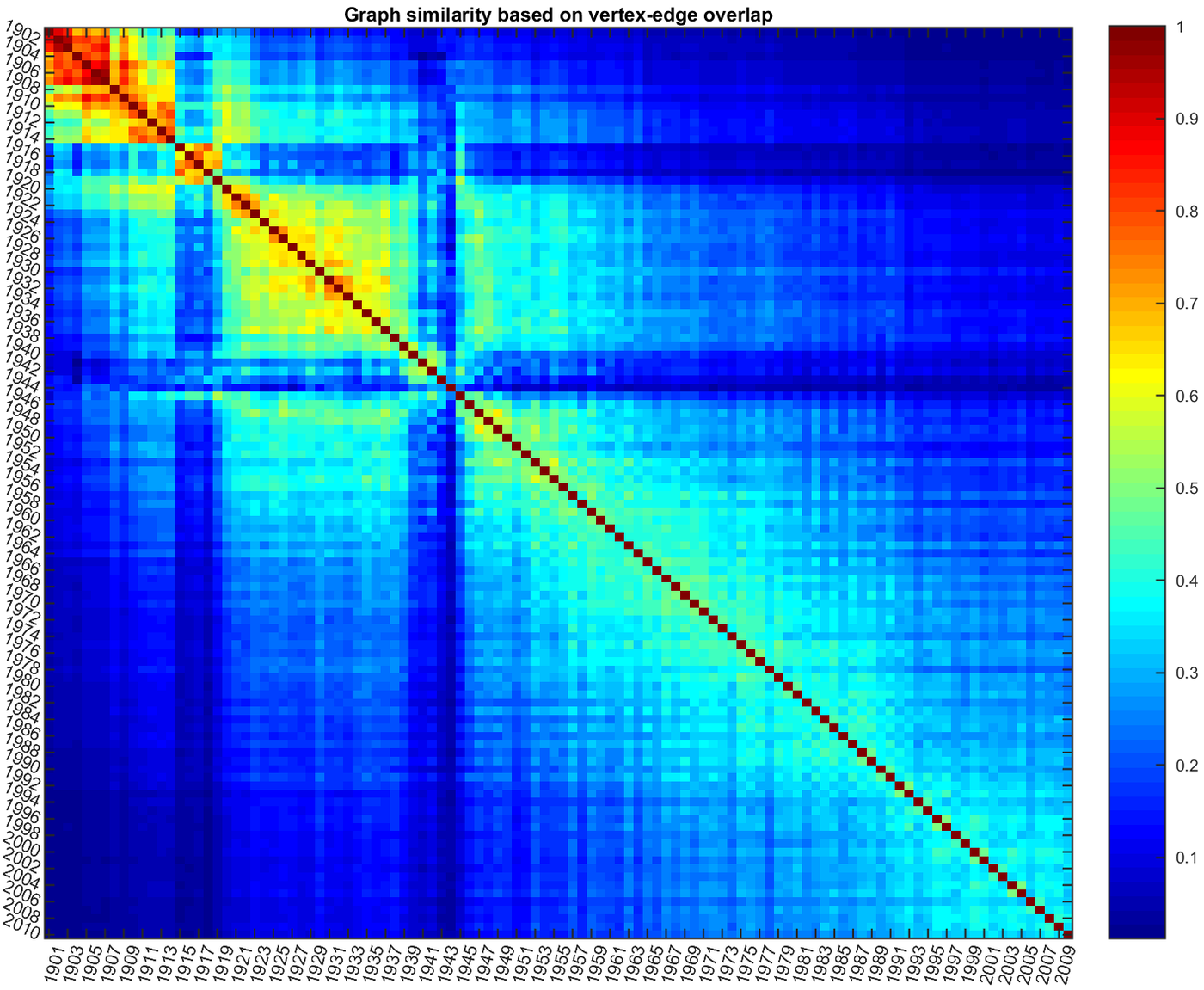}}
\caption{Graph similarity matrices via global graph measures (a) and structural similarity (b)}
\label{fig14:graphlevel}
\end{figure}

Fig.~\ref{fig13:nodelevel} presents the graph similarity matrices obtained based on node-level graph measures (degree, closeness, clustering coefficient, local efficiency). In Fig.~\ref{fig13:nodelevel}, we can see a clear boundary before and after the 1940s due to the second World War, and along the diagonal several blocks could be identified leading to potential temporal states. Fig.~\ref{fig14:a} shows the graph similarity matrix based on global measures. From the image, we can roughly identify two clusters with a blurry boundary around 1950 to 1960. These years could be identified as a transition stage from earlier stage when the football society just started to grow, to current modern stage with established confederations. Fig.~\ref{fig14:b} shows the vertex-edge-overlap similarity matrix, which exhibits a similar pattern with Fig.~\ref{fig13:nodelevel}.

\begin{figure}[h!]
\centering
\includegraphics[scale = 0.7]{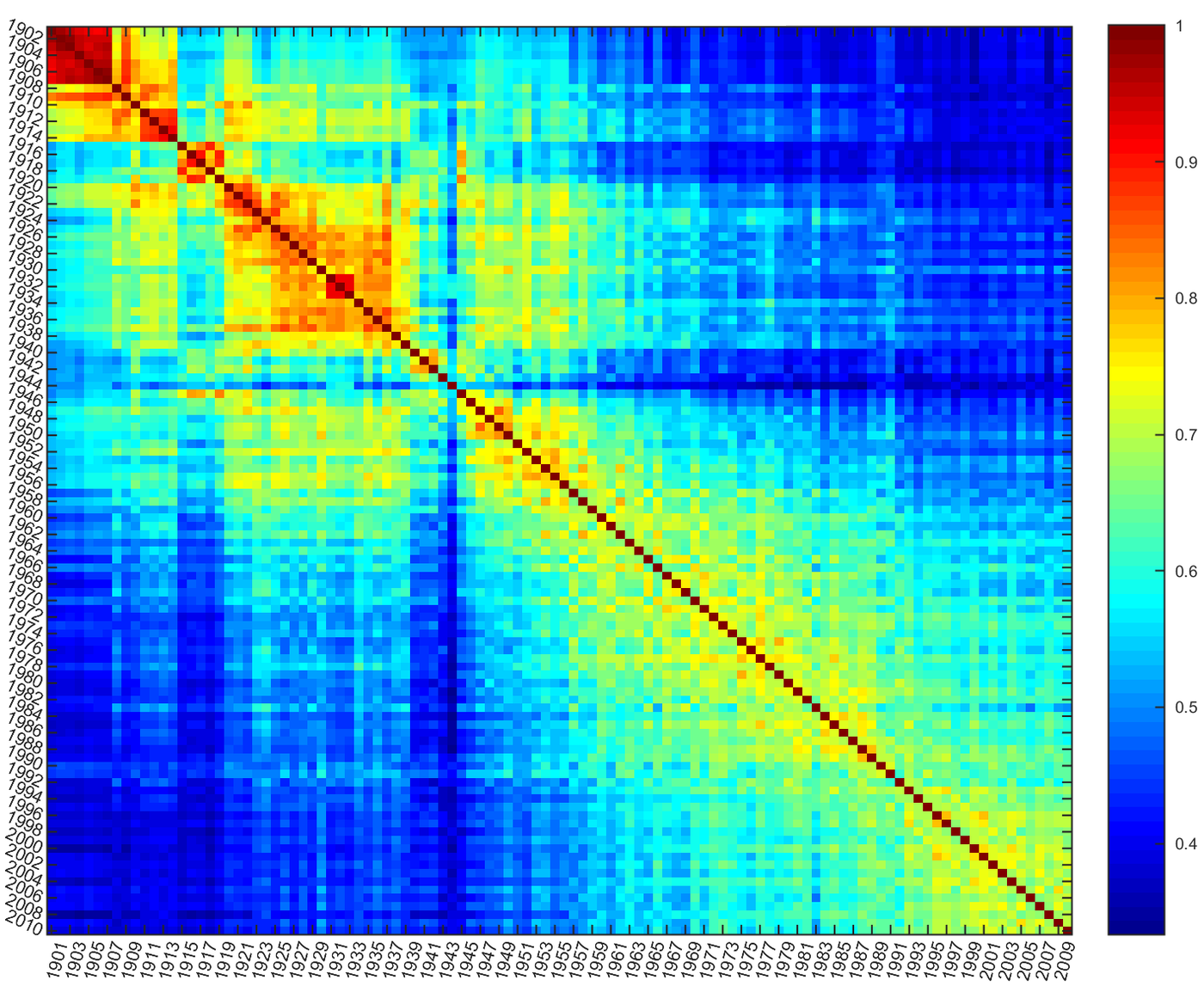} 
\caption{Ultimate graph similarity matrix. The matrix is the average of matrices shown in Fig.~\ref{fig13:nodelevel} and Fig.~\ref{fig14:graphlevel}. The matrix indicates that several states existing in football history.}
\label{fig15:graphsim}
\end{figure}

Fig.~\ref{fig15:graphsim} presents the ultimate graph similarity matrix as the average of all the similarity matrices obtained above. This matrix combines all three levels of similarity, taking features of the network from all aspects into account. We can see clear partitions of the 110 years in football history. If we go along the diagonal line, several individual states could be visually identified. We further carry out a community detection procedure on the similarity matrix of Fig.~\ref{fig15:graphsim}, and partition all the 110 years into community of years, enabling the following temporal states to emerge.

\begin{itemize}
\item[•]\textit{1901-1908}: Start of the history\\
 In early years, only a few countries were playing football, and they played mostly against each other. Football networks in these years share high similarity (see the bright red block on the upper left corner) as they consists of similar nodes and roughly identical edges.
 
\item[•]\textit{1909-1913}: Embryonic form of globalization\\
 In 1908, the first official football tournament was played at the Summer Olympics in London. Most participant countries were from Europe, yet it was the first time that football appeared as an international sport.
 
\item[•]\textit{1914-1918}: WWI\\
The first World War broke out in July 1914 and ended in November 1918. During the war, football in Europe was severely impacted. Fewer international football games were played and some football players even joined the army at that time. Meanwhile, other areas such as South America was less affected. CONMEBOL was founded at this time, and the first Copa América was held in 1916. This explains the high similarity between football networks in these years as shown in Fig.~\ref{fig15:graphsim}.

\item[•]\textit{1919-1938}: Rebuilding\\
After the WWI, peace again returned Europe and football, as the most popular sport there, got prosperous one more time in Europe. Football networks in this stage have high similarities, showing a steady and healthy growth of the game of football. In 1930, Uruguay held the first World Cup with 13 teams. Another 2 World Cups were held in 1934 and 1938 in Italy and France, respectively. Football in this stage showed an increase in the total number of games played, total number of nations participated, and the diversity of the participant nations. Although most nations playing football were from Europe and South America, several Asian and African countries, such as China, Egypt, Palestine, also joined the football world in this period.

\item[•]\textit{1939-1945}: WWII\\
The World War II, from 1939 to 1945, was a disaster to the whole world. In Fig.~\ref{fig15:graphsim}, we can see a blue cross in this time period as the football networks in these years are not at all similar with the rest of the networks. World Cup was forced to stop due to the war. Countries focused merely on how to survive instead of playing football. However, football was not completely forgotten, as in some neutral nations football was still quite popular. Moreover, football was very popular among soldiers and even inside prisoner-of-war camps. 

\item[•]\textit{1946-1959}: Post-war recovery\\
After the second World War, everything began to recover, including football as well. In 1950, Brazil held the 4th World Cup which had been cut off for 12 years. The whole football world started to heal itself. In this period, besides the international football events such as the World Cup and the Olympics, regional football also embraced a fast growth, including
\begin{itemize}
\item[•] UEFA and AFC founded in 1954
\item[•] The first Asian Cup was held in 1956, won by South Korea
\item[•] CAF is founded and the first African Cup of Nations was held in Sudan
\end{itemize}
It is also interesting that graphs in this stage share much similarity with the graphs in the stage before the second World War. This exactly shows the attempt of the football society to get recovered to the status as the one before war.

\item[•]\textit{1960-1990}: Prospering\\
In the previous stage, most regional football confederations were established with the exception of CONCACAF which was founded in 1961. The structure of the modern football world was more or less built up. In this stage, the football world experienced a peaceful growing stage for 30 years. There were not much significant events happening to change the overall landscape. In Fig.~\ref{fig15:graphsim}, we can see a large block from year 1960 to 1990, with a more or less uniform similarity between the networks. 

\item[•]\textit{1991-2010}: The tremendous change\\
In 1991, the Soviet Union dissolved. This historical and political event had its own impact on the football world. In Fig.~\ref{fig15:graphsim}, a new block appears starting from year 1991, indicating that from this moment the network enters a new stage and has less similarity with previous networks. The changes in the new networks are immense as the Soviet Union, which originally was a single and important node in the network, is dissolved into multiple new nodes. With these newly emerged nodes, more football games have been played thus the edges are also significantly influenced. We investigate the political history and generate Table~\ref{t:coldwar}. The nations in the table were either emerged as new nations after the cold war, or got independent from Soviet Union. The table also includes the first football game played by new countries, and the last game before cold war and first game after cold war for other countries. 20 new nodes emerged or re-emerged after Soviet Union dissolved. This political change brings significant alternation of the overall structure of the football network, especially due to the fact that most of the new countries joined UEFA, the confederation with the most impact to the whole football world.

\begin{table}[h!]
\caption{Network nodes emerged after cold war}
      \begin{tabular}{ccc}
        \hline       
        \\[-0.9em]
           Country  & Before cold war   & After cold war\\ \hline
           \\[-0.6em]
           Slovenia & - & 1991\\
           Moldova & - & 1991\\
           Ukraine & - & 1992\\
           Belarus & - & 1992\\
        Uzbekistan &  - & 1992\\
        Kyrgyzstan &  - & 1992\\   
        Turkmenistan &  - & 1992\\
        Tajikistan &  - & 1992\\
        Kazakhstan &  - & 1992\\ 
        Bosnia and Herzegovina & - & 1993\\
        Macedonia & - & 1993\\
        Kosovo & - & 1993\\
        Croatia & 1956 & 1990\\
        Lithuania & 1940 & 1990\\
        Georgia & 1935 & 1991\\
        Latvia & 1942 & 1991\\
        Estonia & 1942 & 1991\\
        Armenia & 1935 & 1992\\
        Azerbaijan & 1935 & 1992\\
        Slovakia & 1944 & 1992\\ 
        \hline
      \end{tabular}
      \label{t:coldwar}
\end{table}

\end{itemize}

The above analysis lists several stages in the football history from year 1901 to year 2010. Over a century of football history is partitioned into 8 states. The interesting fact about these stages is that, although people would assume that the changes of the football networks would be mainly related with the changes within the football world, such as the expansion of World Cup or the commencement of new football tournaments, it is also significantly related with historical and political events, such as wars and political incidents. This shows a perfect evidence of the social impact of football and its ability of offering an insight into the changes in the world. Changes in the world would affect football and football on the other hand, could reflect changes in the world. 

\section*{Discussions and Conclusions}\label{s:conclusion}
This paper analyzes the evolution of football society from a macroscopic point of view. The focus is on the complete football history instead of certain teams, players or leagues as in previous works. Network science disciplines are applied to mine the temporal dynamics and community structures of football networks. Our findings show the existence of community structures in football society, and the identified temporal dynamics demonstrate the continuous growth of the football society with correspondence to globalization process.

Community detection method was later applied on the networks to show the expanding size of communities in each decade. Football communities formed in early years with only a few countries close in geographical distance, and then developed into large ones with nearly all the countries in each continent. The scales of communities are also expanding from local regions to a whole continent. The convergence of the communities and the confederations were also significantly improved from early decades to recent decades.

In this paper we proposed a framework for calculating graph similarity for graph clustering. The framework integrates multiple graph metrics including three levels (node-level, graph-level and structure-level) and considers graph properties from all aspects. Based on the graph similarity matrix, 8 temporal states, each representing one distinct development stage in football history, were found. The temporal states possess great correspondence with both the changes in football society and social events in the world history. 

For the first time, our research analyzes the big picture of the football society and offers a new perspective of the research on football. The method and the framework used in this paper can also be applied on the network analysis on data from other domains. Future research may target at continuous analysis of football data by including more football game records. It would also be a promising aspect to analyze the football network structure using the spectral graph theory and graph signal processing techniques. 


\section*{Declarations}
\begin{backmatter}
\section*{Availability of data and materials}
The datasets supporting the conclusions of this article are available on Eloratings.net. More information of the data can be found at \url{http://www.eloratings.net/}. The datasets can be accessed via \url{https://github.com/YangLiyli131/footballNetworkData}.

\section*{Competing interests}
  The authors declare that they have no competing interests.
  
\section*{Funding}
  Work in this paper was supported by the NSF award CCF-1750428.
  
\section*{Author's contributions}
YL processed and analyzed the data and conducted the research scheme in this work. YL was responsible for generating the plots and tables in this work. GM provided supervision of the whole research plan and contributed to the polishing of this work. All authors read and approved the final manuscript.
  
\section*{Authors' information}
YL received a B.Sc. degree in Electrical Engineering (Automation) from Tianjin University, Tianjin, China, in 2013, and a M.Sc. degree in Electrical and Computer Engineering from the University of Rochester in 2016. He joined the Ph.D. program at the University of Rochester in January 2017.

GM is an Associate Professor with the Dept. of Electrical and Computer Engineering, University of Rochester as well as an Asaro Biggar Family Fellow in Data Science. He is also affiliated with the Goergen Institute for Data Science.

\section*{Acknowledgements}
  Not applicable.
\end{backmatter}



\bibliographystyle{bmc-mathphys} 
\bibliography{paper}      
\nocite{label}



\end{document}